\documentclass[aip,jcp,amsmath,amssymb,reprint]{revtex4-1}

\usepackage[utf8]{inputenc}
\usepackage[T1]{fontenc}
\usepackage{etoolbox}
\usepackage{graphicx}
\usepackage{dcolumn}
\usepackage{bm}
\usepackage{blindtext}
\usepackage{physics}
\usepackage{siunitx}
\usepackage{float}
\usepackage{xcolor}
\usepackage{bbold}
\usepackage{relsize}
\usepackage[acronym]{glossaries}
\usepackage[normalem]{ulem}
\usepackage{listings}
\usepackage{pythonhighlight}

\newcommand{\ttiny}[1]{\text{\tiny{#1}}}

\definecolor{UniBlue}{HTML}{5F6DA3}
\definecolor{TitleBlue}{HTML}{EBEDF5}
\definecolor{TxtBlue}{HTML}{182D80}
\definecolor{TxtRed}{HTML}{CC3300}
\definecolor{TxtGreen}{HTML}{339900}
\definecolor{SecBlue}{HTML}{434E76}
\definecolor{TxtBlack}{HTML}{212121}
\definecolor{TxtGrey}{HTML}{828282}

\begin{document}


\title[]{FeNNol: an Efficient and Flexible Library for Building Force-field-enhanced Neural Network Potentials}

\author{Thomas Plé}
\email[Corresponding author: ]{thomas.ple@sorbonne-universite.fr}
\affiliation{Sorbonne Université, LCT, UMR 7616 CNRS, 75005, Paris, France}
\author{Olivier Adjoua}
\affiliation{Sorbonne Université, LCT, UMR 7616 CNRS, 75005, Paris, France}
\author{Louis Lagardère}
\affiliation{Sorbonne Université, LCT, UMR 7616 CNRS, 75005, Paris, France}
\author{Jean-Philip Piquemal}
\email[Corresponding author: ]{jean-philip.piquemal@sorbonne-universite.fr}
\affiliation{Sorbonne Université, LCT, UMR 7616 CNRS, 75005, Paris, France}


\date{\today}

\begin{abstract}
Neural network interatomic potentials (NNPs) have recently proven to be powerful tools to accurately model complex molecular systems while bypassing the high numerical cost of \textit{ab-initio} molecular dynamics simulations. In recent years, numerous advances in model architectures as well as the development of hybrid models combining machine-learning (ML) with more traditional, physically-motivated, force-field interactions have considerably increased the design space of ML potentials. In this paper, we present FeNNol, a new library for building, training and running force-field-enhanced neural network potentials. It provides a flexible and modular system
for building hybrid models, allowing to easily combine state-of-the-art embeddings with ML-parameterized physical interaction terms without the need for explicit programming. Furthermore, FeNNol leverages the automatic differentiation and just-in-time compilation features of the Jax Python library to enable fast evaluation of NNPs, shrinking the performance gap between ML potentials and standard force-fields. This is demonstrated with the popular ANI-2x model reaching simulation speeds nearly on par with the AMOEBA polarizable force-field on commodity GPUs (GPU=Graphics processing unit). We hope that FeNNol will facilitate the development and application of new hybrid NNP architectures for a wide range of molecular simulation problems.

\end{abstract}

\maketitle


\section{Introduction}

The field of large-scale molecular dynamics (MD), has traditionally relied on the use of force fields (FFs) to describe interatomic interactions. Force fields can leverage different functional forms, from simple classical fixed charge models such as AMBER~\cite{wang2004development}, CHARMM~\cite{vanommeslaeghe2010charmm} or OPLS \cite{OPLS} to the more \textit{ab-initio}-inspired ones such as SIBFA~\cite{gresh2007anisotropic,naseem2022development}, GEM~\cite{piquemal2006towards,duke2014gem} or MB-pol~\cite{reddy2016accuracy,zhu2023mb}, with a spectrum of intermediate ones such as AMOEBA~\cite{AMOEBA03,ponder2010current}, AMOEBA+ \cite{AMOEBA+1,AMOEBA+2} or ARROW~\cite{nawrocki2022protein}. Each form strikes a unique balance between representational power and computational efficiency and the choice between force fields depends on the complexity of the system under study,\cite{melcr2019accurate, reviewcompchem,annurev-biophys-070317-033349} as well as the availability of model parameters. Force-fields have reached a high maturity over the years, with automatic parameterization tools~\cite{wang2001antechamber,walker2022automation,wang2022end,chen2023advancing} and efficient implementations~\cite{thompson2022lammps,case2008amber,van2005gromacs,phillips2005scalable,lagardere2018tinker,adjoua2021tinker}, enabling their use for larger and more complex systems. Despite continuous refinements, traditional FFs have inherent limitations due to their fixed functional form which often relies on \textit{ad-hoc} atom types, and to the complexity of the parameterization task.

In recent years, a new breed of potentials has emerged, powered by machine learning (ML) techniques. These ML potentials, including for example neural network interatomic potentials (NNPs), leverage flexible functional forms and large amounts of \textit{ab initio} data to reach very high accuracy on broad classes of systems. This makes them a promising alternative to traditional FFs, especially for modeling exotic systems or chemical reactivity. Over the last few years, a large variety of model architectures have been developed, using various computational frameworks such as kernel models~\cite{chmiela2018towards,bigi2023wigner}, NNPs with fixed-form symmetry functions~\cite{shakouri2017accurate,smith2017ani}, message-passing neural networks~\cite{schutt2017schnet,lubbers2018hierarchical,zubatyuk2019accurate}, E(3)-equivariant tensor computations~\cite{qiao2021unite,unke2021spookynet,gasteiger2021gemnet,batzner20223,musaelian2022learning} or multi-scale representations~\cite{grisafi2019incorporating,grisafi2021multi}. The development of hybrid models, which combine ML with more traditional, physically-motivated force-field interactions, has further expanded the design space of ML potentials~\cite{ko2021fourth,unke2019physnet,unke2021spookynet,qu2023interfacing,yang2022transferrable,zhang2022deep,bowman2022delta,zhu2023mb,chen2023phyneo,AMOEBA-NN,ple2023fennix}.

However, the advent of these new models and architectures has brought its own set of challenges. Prominently, numerical implementations of these architectures are currently scattered in many packages provided by the authors of the methods~\cite{}. This makes a direct comparison between the models a difficult task, in particular with respect to computational efficiency. A few existing libraries provide a wider scope and more modular implementations such as SchnetPack~\cite{schutt2023schnetpack}, MLAtom~\cite{dral2024mlatom} or deepmd-kit~\cite{zeng2023deepmd} but they are still limited in the range of models offered to the user. Furthermore, designing models for non-standard workflows usually requires custom implementations, thus imposing barriers at entry for non-expert users. Examples of such applications include, for example, fitting distributed multipoles~\cite{bereau2015transferable,glick2021cartesian,thurlemann2022learning} or a decomposition of the energy, or feeding force-field terms with ML-generated parameters, as was done in our previously proposed FeNNix model~\cite{ple2023fennix}.

To address these challenges, we present FeNNol, a new Python library designed for building, training, and running atomistic machine-learning models, with a particular focus on physics-enhanced neural network potentials. FeNNol provides a flexible and modular system that allows users to easily build custom models, allowing for example the combination of state-of-the-art atomic embeddings with ML-parameterized physical interaction terms, without the need for explicit programming. FeNNol leverages the Jax framework~\cite{jax2018github,deepmind2020jax} and its just-in-time compilation capabilities, and provides a collection of efficient and configurable modules that can be composed to form complex models.

The paper is structured as follows: in section~\ref{section:architecture}, we describe FeNNol's software architecture, detailing its design principles and the underlying technologies it leverages. We focus in particular on FeNNol "modules" which are at the core of model building. In section~\ref{section:crate}, we introduce a novel multi-paradigm atomic embedding which serves as a demonstration of the highly configurable nature of FeNNol modules. In section~\ref{section:training}, we describe FeNNol's training system that enables users to define complex models and train them on generic tasks. In section~\ref{section:md}, we present the different ways to use FeNNol models in molecular dynamics simulation. We then demonstrate the performance of FeNNol's models and native MD engine by showing that our implementation of the popular ANI-2x model~\cite{devereux2020extending} reaches simulation speeds close to the optimized GPU-accelerated (GPU=Graphics Processing Unit) Tinker-HP implementation of the AMOEBA force-field~\cite{adjoua2021tinker}. We also showcase our interface with Deep-HP~\cite{inizan2022scalable} that enables efficient NNP simulations of systems reaching half a million atoms. Finally, in section~\ref{section:examples}, we provide concrete examples of model training with FeNNol, demonstrating how one can easily build non-standard models and how state-of-the-art results can be achieved within a few hours of training on a commodity GPU.

\section{Software Architecture}\label{section:architecture}
FeNNol is built using the Jax library~\cite{jax2018github} that enables developers to easily write GPU-accelerated and automatically differentiable Python functions. More specifically, it relies on the Flax ML framework~\cite{flax2020github} to build parameterized models (such as neural networks) in a compact, highly configurable and modular way. Model building in FeNNol is organized around Flax Modules that each implement a specific functionality (for example computing an atomistic embedding, aggregating pairwise information, evaluating a neural network, etc...). By convention, modules take as input a dictionary containing information on the molecular system (coordinates, species, batch indices, etc...) and all information provided by previous modules. Each module thus uses this information to perform its computation and returns a new enriched dictionary that is passed to subsequent modules. In the following, we shall refer to this dictionary as the \textit{system state} (as opposed to the \textit{model state} that contains the different parameters and mutable state variables of the model).

An important feature of Jax is its ability to just-in-time (JIT) compile and optimize code (using the XLA compiler), thus merging modules and removing unused code at execution. This allows for very configurable modules while retaining efficiency at runtime. The principal drawback of this approach (compared to eager execution in Pytorch for example) is that all array sizes must be known at compilation time. The main restriction that this imposes is for the construction of neighbour lists (the list of pairs of atoms with distance under a certain cutoff) which size may be unknown in advance. The construction of neighbour lists thus cannot be fully compiled and we resolve this issue by using a preprocessing pipeline that is not fully JIT compiled nor differentiated (details in section~\ref{sec:preprocessing}).\\

\begin{figure*}
    \centering
    \includegraphics[width=1.\textwidth]{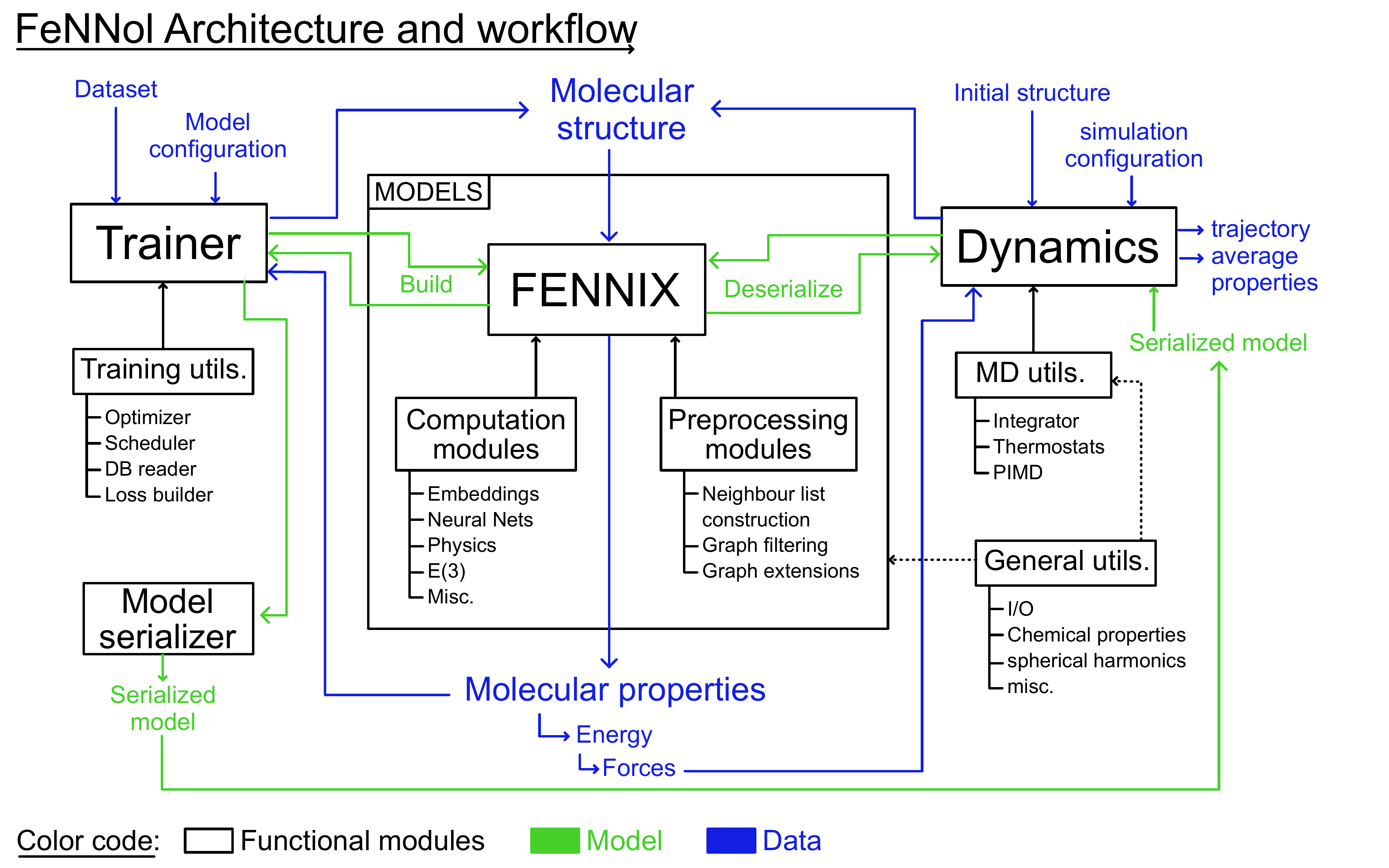}
    \caption{Architecture and data flow of the FeNNol library. Black blocks represent functional modules, green arrows represent model operations and blue arrows represent data flows.}
    \label{fig:architecture}
\end{figure*}

A global description of the software architecture and data flows is provided in figure~\ref{fig:architecture}. In the following subsections, we describe the different modules available to build models, starting with the preprocessing modules that handle operations on neighbour lists and following with the computation modules that handle the bulk of models' logic and are divided in a few categories: atomistic embeddings, chemical and radial encodings, physics modules, neural networks and operation modules. We will then describe how one can extend FeNNol by registering custom modules. Finally, we provide an example of how to build a full model with the \verb|FENNIX| constructor which is the main entry point for model building.

\subsection{Preprocessing Modules}\label{sec:preprocessing}
As explained earlier, a preprocessing pipeline that is not compiled is necessary to build neighbour lists without assuming a maximum size in advance. The preprocessing modules thus handle operations on neighbour lists -- or more generally molecular graphs -- such as building, filtering, padding or extending.\\

The main preprocessing module is the \verb|GraphGenerator| which constructs a molecular graph given a cutoff distance, a set of atom coordinates, and optionally cell vectors for periodic boundary conditions. To avoid padding the array of coordinates when a batch of different molecular structures is provided, we chose to handle a concatenated array of all the atoms, accompanied by an array keeping track of the batch index of each atom. A molecular graph consists \textit{a minima} of two arrays of atom indices: the source and destination of each edge in the molecular graph. An edge is assigned between two atoms when they have the same batch index and the distance between them is lower than the cutoff distance. Note that in the current implementations, all graphs are directed (meaning that each pair of atoms appears twice in the neighbour list, with source and destination atoms reversed). If cell vectors are provided, edges are assigned for each pair of images within the cutoff distance (with at least one of the atom in the main periodic cell) or according to the minimum image convention if required by the user. When periodic boundary conditions (PBCs) are applied, an array containing the shift (in units of cell vectors) to apply to the vector connecting the source and destination atoms is added to the graph. 
In the current implementation, neighbour lists are obtained by first computing all distances between pairs of atoms (that have the same batch index) and then filtering to the cutoff distance. This implementation is efficient for small number of atoms and, with GPU acceleration (and using a neighbour list skin in MD simulations to compute all distances less often), we found that systems up 20,000 atoms do not suffer too much from the quadratic scaling. Future releases will use cell lists to improve the scaling of neighbour lists construction.
Finally, when the model uses long-range interactions that require Ewald summations\cite{ewald1921ewald}, the user can configure the \verb|GraphGenerator| module to construct vectors from the reciprocal lattice and filter them according to a magnitude cutoff.

In some cases, it can be useful to filter the molecular graph according to a shorter cutoff distance. This would for example happen when long-range physical interaction (for example dispersion interactions) are coupled to a shorter-range ML embedding. In that case, the \verb|GraphGenerator| module would construct the longer-range graph and the \verb|GraphFilter| module can be used to select only the pairs within the shorter cutoff. This outputs a separate graph (which is marked as a subgraph of the original longer-range graph) with the filtered source and destination arrays and a new array containing a mapping of edges in the new graph to indices in the original graph. This map can be used by subsequent modules to transfer edge properties between the two graphs. 

Finally, we provide the \verb|GraphAngularExtension| module that builds a list of pairs of edges with a common central atom. This is used when explicit three-body terms are computed by the model, for example in Behler-Parrinello~\cite{behler2007generalized} or ANI~\cite{smith2017ani} symmetry functions. This module does not produce a new graph but simply enriches the input graph.\\

In order to minimize re-compilation of the main modules, the size of neighbour lists must not change often. Thus, each array which size is the number of edges in the graph is padded to the largest neighbour list size encountered up to the current execution. A state variable thus keeps track of this maximum size which is updated to the current list size if it exceeds it (with an optional multiplicative factor to avoid too frequent re-compilations). In practice, for example when performing molecular dynamics, this implies multiple re-compilation in the beginning of the simulation. When the maximum size of the neighbour list is found, the main modules do not need to re-compile and the simulation achieves its full efficiency. This can be achieved faster by increasing the multiplicative factor for the padding. The additional edges from the padding however impose an overhead on the computation and a balance between fast convergence of the list size and minimal padding must be found for the specific model and simulated system.\\

Finally, since, the preprocessing modules are not differentiated, the distances computed in these modules cannot be used to compute forces (or coordinates-dependent derivatives in general) as they do not explicitly depend on the coordinates array passed to the main modules. Thus, for each preprocessing module, a matching \textit{processing} module is automatically added to the list of main modules when constructing a full model (see section~\ref{sec:building_model}). These processing modules compute edge vectors and distances with the provided coordinates and neighbour lists, as well as angles between pairs of edges if necessary. Furthermore, they compute a switching function that depends on the distance and goes smoothly to zero as it approaches the cutoff distance. This switching function is used by subsequent modules to avoid discontinuities in molecular properties when atoms enter or leave an atomic neighbourhood. The switching function can be configured by the user with optional trainable parameters. Functions available in the current implementation are the cosine function (as used in the ANI~\cite{smith2017ani} models for example), the polynomial envelope from ref.~\onlinecite{gasteiger2020directional} and the exponential function from SpookyNet~\cite{unke2021spookynet}.

\subsection{Computation Modules}
In this section, we describe the main modules responsible for the bulk of the computation in a model. A model is constituted by a chain of such modules that perform operations on the molecular graphs and system state such as computing an atomic environment embedding (or descriptor) or processing it with a multi-layer perceptron to compute the energy. This chain of modules is encapsulated in a top-level Flax module which is JIT-compiled and acts as a pure function that can be automatically differentiated either with respect to model parameters (for training) or to inputs (for example to compute forces). In the following, we list the different categories of modules and provide some examples.\\

\paragraph{Atomic Embeddings}~\\
Atomic embedding modules process the molecular graph to compute descriptors of the atoms' environments that can later be used by other modules (such as neural networks) to compute properties of the system (for example the energy). Embeddings are usually of atom-wise or pair-wise nature and can be either local (that only process directly connected atoms), message-passing (that capture longer-range interactions via iterative exchanges of information with neighbouring atoms) or completely non-local (for example global embeddings such as Coulomb matrices~\cite{rupp2012fast} or non-local atom-wise like in the LODE framework~\cite{grisafi2019incorporating,grisafi2021multi} or the attention mechanism in SpookyNet~\cite{unke2021spookynet}). FeNNol provides various embeddings from the literature: local models such as ANI AEVs~\cite{smith2017ani}, Allegro~\cite{musaelian2022learning}, EEACSF~\cite{eckhoff2023lifelong} and DeepPot-SE~\cite{zhang2018end} and message-passing models such as AIMNet~\cite{zubatyuk2019accurate}, HIP-NN~\cite{lubbers2018hierarchical}, PaiNN~\cite{schutt2021equivariant} and NewtonNet~\cite{haghighatlari2022newtonnet}. When necessary, E(3)-equivariant operations (such as tensor products) are performed using the \verb|e3nn-jax|~\cite{geiger2022e3nn} library. In some simple cases, we implemented equivariant operations from scratch to increase performance (for example, the filtered tensor product proposed in ref.~\onlinecite{kozinsky2023scaling}). We also implemented a new multi-paradigm embedding denoted as CRATE that will be described in section~\ref{section:crate}.\\

\paragraph{Chemical and Radial Encodings}~\\
Basic building blocks of most atomic embeddings are chemical and radial encodings that capture information about the chemical species of each atom and the distance between pairs of atoms. 

The chemical encoding maps each chemical species to a multidimensional vector that can be either fixed or learned during training and that will be further processed by the atomic embedding to combine it with geometric information. Different chemical encodings will provide different prior knowledge to the model about the relations between chemical species. For example, a simple one-hot encoding makes no particular assumptions while the positional encoding of ref.~\onlinecite{ple2023fennix} or the electronic structure encoding of ref.~\onlinecite{unke2021spookynet} assume similarities for atoms close together in the periodic table. FeNNol provides a few different chemical encodings: the one-hot encoding, an encoding based on electronic occupancy as introduced in TeaNet~\cite{takamoto2022teanet}, the electronic structure encoding from~\onlinecite{unke2021spookynet}, the positional encoding from ref.~\onlinecite{ple2023fennix}, a 4-dimensional encoding based on the Stowe-Janet-Scerri periodic table. The encoding can also be initialized as a random vector for each species. These encodings can be combined together via concatenation and optionally be optimized during training.

The radial encoding is a projection of the distance between two atoms on a set of smooth basis functions. FeNNol implements multiple families of basis functions: the simple Gaussian basis used for example in the ANI models~\cite{smith2017ani}, the Bessel basis introduced in ref.~\onlinecite{gasteiger2020directional}, the inverse-distance Gaussian used in HIP-NN~\cite{lubbers2018hierarchical}, a Fourier expansion and the Bernstein polynomials used in ref.~\onlinecite{unke2021spookynet}. As the nature and number of basis functions directly influence the spatial regularity of the model, it can drastically impact the fit of target properties so that comparison between different radial basis can be useful when designing a model. \\

\paragraph{Physics Modules}~\\
In order to build hybrid models that incorporate long-range effects such as FENNIX-OP1~\cite{ple2023fennix}, FeNNol offers multiple physically-motivated interaction modules. It provides for example a Coulomb module that computes electrostatic interactions between distributed charges with different short-range damping schemes available. In this module, long-range interactions in periodic systems are handled using Ewald summations (later versions of FeNNol will include an implementation of Smooth Particle Mesh Ewald\cite{essmann1995smooth} for more efficient interactions).
Other available physics modules include the recently proposed QDO dispersion and exchange interactions~\cite{khabibrakhmanov2023universal}, charge equilibration (as defined in DFT-D4~\cite{caldeweyher2019generally}) and the ZBL repulsive pair-potential~\cite{ziegler1985stopping}.
Most of the physics modules can be parameterized using dynamically computed properties (for example obtained via a neural network) as is done for example with atomic volumes parameterizing electrostatics and dispersion interaction in ref.~\onlinecite{ple2023fennix}. \\

\paragraph{Neural Networks}~\\
The main use of neural networks in atomistic models is to process embeddings and produce outputs (for example atomic energies, charges, etc...). The neural network weights can then be fitted against relevant targets in a dataset or used to parameterize physical interactions. FeNNol provides multiple neural network architectures: a simple multi-layer perceptron, a residual neural network and various chemical-species-dependent neural networks. For the latter, different levels of species specialization are provided: full species-dependent network parameters, shared parameters with species-dependent activation scales and bias or low-rank chemical adaptation (ZLoRaNet) inspired from the LoRa methodology of fine-tuning~\cite{hu2021lora}. \\

\paragraph{Operation Modules}~\\
Finally, we implemented various miscellaneous modules that perform simple operations on the system state such as reshaping an entry, adding multiple entries together, sum bond-wise properties to atomic centers or apply various shifts, scaling and activation functions. These can be used to construct more complex models from simple building blocks directly in the model input instead of coding explicit custom modules. A basic example would be to use the \verb|Concatenate| module to add charge information to a short-range embedding as is done in 4G-HDNNP~\cite{ko2021fourth}. This flexibility aims at letting the user quickly experiment and iterate with the model architecture. \\

\subsection{Extensions with custom modules}
As mentioned in the beginning of section~\ref{section:architecture}, a FeNNol module is simply a Flax module (i.e. a class inheriting from \verb|flax.linen.Module|) which \verb|__call__| method takes the system state (the dictionary containing all previously computed properties of the system) as input and returns a new processed system state as output. As an example, the following code shows the \verb|Activation| module :\\

\begin{verbatim}
class Activation(flax.linen.Module):
  key: str
  activation: Union[Callable, str] = nn.silu
  scale_out: float = 1.0
  shift_out: float = 0.0
  output_key: Optional[str] = None

  ### FeNNol ID for automatic registration
  FID: str = "ACTIVATION"

  @flax.linen.compact
  def __call__(self, inputs:Dict[str,Any]):

    ## get the entry to which we want 
    ## to apply the activation function
    x = inputs[self.key]
 
    ## get activation function
    activation = (
        activation_from_str(self.activation)
        if isinstance(self.activation, str)
        else self.activation
    )
 
    ## apply activation and
    ##   linear transformation
    output =  (
      self.scale_out * activation(x)
       + self.shift_out
    )
 
    ## determine output key
    if  self.output_key is None:
      output_key = self.key
    else:
      output_key = self.output_key
 
    ## return new system state
    return {**inputs, output_key: output}
\end{verbatim}  
~\\
This module applies an activation function followed by a linear transformation to an entry in the system state. As flax modules are frozen dataclasses, all their attributes are specified and explicitly typed in the class definition, the class constructor is automatically built and attributes are immutable after initialization (see the Flax documentation for more details). Note that for a flax module to behave properly, the \verb|__call__| method must be a pure function. This is why we do not explicitly modify the \verb|inputs| dictionary but rather return a shallow copy with additional keys. A key advantage of the jax/flax framework is that all non-jax operations are considered static by the jit compiler and thus executed only once at compile-time. This allows modules to be extremely configurable without imposing computational overheads. For example, the \verb|activation_from_str| function in the previous code example parses the \verb|self.activation| attribute to build an activation function from a character string. By construction, the attributes of flax modules are immutable so that \verb|self.activation| is guaranteed to never change and it would thus be wasteful to parse it again at each call of the model. Thanks to the jit system, this parsing is only done once at compile-time. \\

When importing the FeNNol library, all the flax modules in the source code with the \verb|FID| (Fennol IDentifier) attribute are automatically registered as FeNNol modules and available when building a model. FeNNol also automatically registers modules from the Python files found in repositories listed in the environment variable \verb|FENNOL_MODULES_PATH|. The user can also manually register a module after initialization by providing their custom module and an FID to the \verb|register_fennix_module| function before building or loading a model.

\subsection{Building a FeNNix model}~\label{sec:building_model}
In FeNNol, a model is constructed by calling the \verb|FENNIX| constructor which most important arguments are two ordered dictionaries: one that lists the preprocessing modules and their properties and another one for the computation modules with their properties:
\begin{verbatim}
    from fennol import FENNIX
    import jax
    rng_key = jax.random.PRNGKey(123456)
    
    model = FENNIX(
              cutoff=5.,
              modules={...},
              preprocessing={...},
              rng_key=rng_key,
              energy_terms=["energy"],
            )
\end{verbatim}
The model automatically builds a primary graph with a cutoff radius specified by the \verb|cutoff| argument. This primary graph can optionally be customized using the preprocessing parameters. The argument \verb|rng_key| is a key for the random number generator that is used to initialize the model parameters. The \verb|energy_terms| argument defines which keys in the final system state will be used as energy components and summed to form the total energy.
If only one graph is required (i.e. no filtering and no angle information needed), the preprocessing argument can be omitted. The \verb|modules| dictionary has the following structure:
\begin{verbatim}
   modules = {
      "module1":{
         "FID": MODULE_1_FID
         "param1": ...
         "param2": ...
      },
      "module2":{
        ...
      },
      ...
    }
\end{verbatim}
The key for each module will be used as the \verb|name| attribute of the underlying flax module and may be used to setup output keys. The \verb|"FID"| key is the identifier of the module. Finally, the remaining entries set values for the module parameters (i.e. the dataclass attributes). A list of registered modules can be obtained by calling the \verb|available_fennix_modules| function. 

After manipulations (for example training), a model can be serialized and written to a file using the \verb|FENNIX.save| method as 
\begin{verbatim}
    model.save("model_name.fnx")
\end{verbatim}
By convention, we use the \verb|.fnx| extension to identify model files. This file contains all the information needed to reconstruct the model. This can be achieved with the \verb|FENNIX.load| class method as 
\begin{verbatim}
    model = FENNIX.load("model_name.fnx")
\end{verbatim}

\begin{figure}
    \centering
    \includegraphics[width=0.45\textwidth]{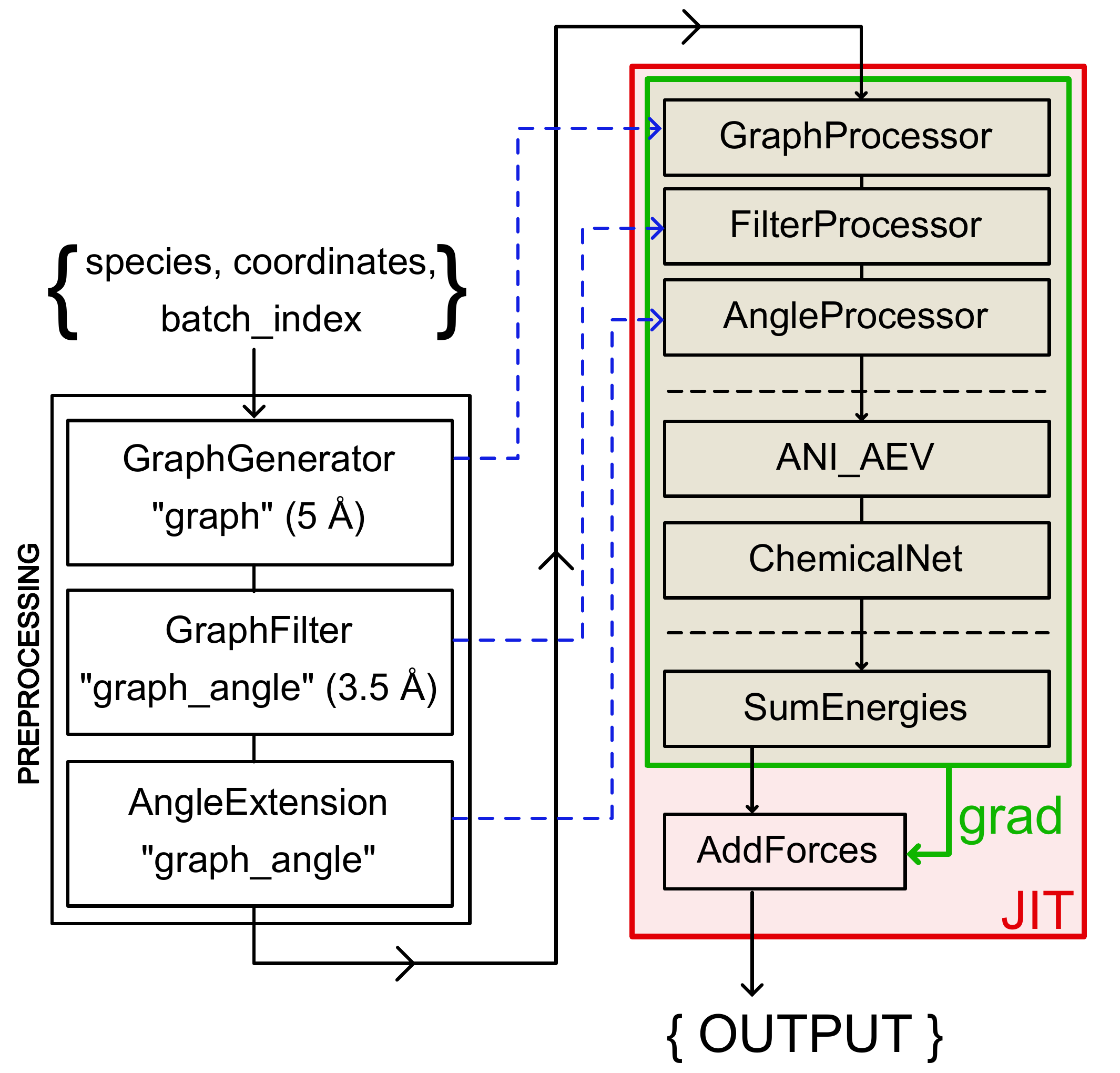}
    \caption{Flow diagram of the execution of a ANI-like model built with FeNNol}
    \label{fig:ani_diagram}
\end{figure}

A flow diagram of the execution of a ANI-like model built with FeNNol is provided in figure~\ref{fig:ani_diagram}. The input is built from chemical species (atomic numbers) and coordinates. In case where multiple structures are provided (batching), all the atoms are concatenated in the \verb|species| and \verb|coordinates| arrays and an additional \verb|natoms| array is provided containing the number of atoms of each structure in the batch and is used to compute the batch index of each atom. The first modules on the left are the preprocessing modules that build the neighbour list (with a cutoff of 5 \AA), filter it to a new graph with a cutoff of 3.5 \AA~ and finally extend this graph with indices of pairs of neighbours forming angles that will be used to construct the environment vector in the \verb|ANI_AEV| module. The blue arrows indicate graph processing modules that are automatically added by the corresponding preprocessing modules. As explained earlier, these processing modules compute distances and angles using provided edge indices that will be automatically differentiated (green box in the diagram) to obtain atomic forces. The next two modules are the user-defined main modules (here \verb|ANI_AEV| that computes the atomic environment vectors of the ANI model~\cite{smith2017ani} and \verb|ChemicalNet| that applies a species-specialized neural network to the AEVs to obtain atomic energies). The last automatically differentiated block simply sums the energy contributions of each atom to the corresponding structure's total energy. The final module adds atomic forces obtained from the \verb|jax.grad| transform to the output.

\section{The "CRATE" multi-paradigm embedding}\label{section:crate}
In order to exploit the highly configurable nature of flax modules, we developed the CRATE (Configurable Resources from ATomic Environment) atomic embedding. The idea behind CRATE is to combine chemical and geometric information information from different sources: chemical information from species embeddings, radial and angular information from radial basis and angle embeddings, tensorial information from E(3)-equivariant operations and long-range information using the LODE methodology. The user can specify which resources they need, depending on the nature of the system, their fitting targets (for example fitting distributed multipoles would require tensorial information while energies might only require radial and angular information) and their computational budget. These resources can also be iteratively combined via interaction layers using message-passing or, if locality is important, using simulated message-passing with a local key-query mechanism that we detail in Appendix~\ref{appendix:crate} or with directed edge embeddings. Finally, we designed the CRATE components to minimize expensive operations on graph edges or triplets. For examples, neural networks applied to high-dimensional vectors or expensive E(3) tensor products are only performed on atomic centers. Edge or triplet operations then mostly consist in element-wise addition, multiplication or function activation, as well as (generalized) outer products. A detailed description of the model architecture and the composition of the different resources can be found in Appendix~\ref{appendix:crate}.

By selecting specific resources and interaction layers, the CRATE embedding can be made to operate similarly to well-established architectures. For example, selecting radial and angular information with only one interaction layer, the model behaves like the ANI or BPNN architectures. Keeping the same resources and adding multiple message-passing layers leads to an architecture that resembles AIMNet~\cite{zubatyuk2019accurate}. Finally, if we select E(3)-equivariant resources, refine them with local atom-wise tensor products and combine them using message-passing interactions, we obtain a model close to the recently proposed MACE~\cite{batatia2022mace}. Each interaction layer can be configured separately so that many combinations are possible and one can tailor the architecture for their data and computational efficiency requirements.

\section{Model Training}\label{section:training}
FeNNol provides a basic training system that should handle most simple model training tasks. This training system can be invoked with the command \verb|fennol_train| that is installed with the FeNNol Python package. The input to the training program is a structured file (either yaml, json or toml) that describes both the training parameters and the model structure. The latter is the same as presented in section~\ref{sec:building_model}. The training parameters can be subdivided in three categories: general parameters, optimizer and schedule parameters and loss function definition.\\

\paragraph*{General Parameters} General parameters include the path to the dataset, the batch size, number of batches per epoch, maximum number of epochs, the names of the output directory and log file, seed for the random number generator, etc... \\

\paragraph*{Optimizer and schedule parameters} Model optimization is performed via stochastic gradient descent using optimizers provided by the Optax library~\cite{deepmind2020jax}. The default optimizer is the Adabelief~\cite{zhuang2020adabelief} which we found to perform slightly better than the more common Adam optimizer. The optimizer can be chosen and configured by the user by providing the name of the Optax optimizer and its parameters, as described in the Optax documentation. Decoupled weight decay~\cite{loshchilov2017decoupled} can be applied (independently of the chosen optimizer) to all or a subset of user-selected parameters. Additionally, parameter gradients can be clipped using the adaptive gradient clipping methodology~\cite{brock2021high}. Finally, the user can define a subset of modules and/or parameters to be frozen, meaning that their parameters will not be updated (this is useful for example in a fine-tuning or transfer-learning stage).\\
The learning-rate scheduler can also be configured by the user with a choice between a constant learning-rate, a cosine-one-cycle schedule~\cite{smith2019super} or an adaptive schedule similar to Pytorch's \verb|reduce_lr_on_plateau|.\\

\paragraph*{Loss definition} The full loss function for the model optimization is defined in the training configuration file. This definition takes the form of a dictionary of loss function components with parameters such as the reference key in the dataset, the key to use in the model output, the weight of the loss component and the type of loss function to use. The following listing (in yaml format) provides an example for a mean square error loss function on both energy and forces:
\begin{verbatim}
  loss:
    e:                        
      key: total_energy     
      ref: formation_energy   
      weight: 1.e-3  
      type: mse
      per_atom: True
    f:
      key: forces
      ref: forces
      weight: 1.
      type: mse
\end{verbatim}
Multiple loss types are provided such as means squared error, mean absolute error, log hyperbolic cosine, negative log-likelihood or CRPS for ensemble training~\cite{kellner2024uncertainty} or evidential losses~\cite{amini2020deep,meinert2021multivariate}. The loss type can be set for all components at once using the \verb|default_loss_type| parameter. The \verb|per_atom| directive only applies to system-wise properties and divides the loss by the number of atoms. This can be useful when training on systems of widely varying sizes.\\

\paragraph*{Training stages} When training a model on a multi-component loss function, it may be useful to define a multi-stage training. As a concrete example, we can mention the training of the FENNIX-OP1 model~\cite{ple2023fennix} which required to first train on short-range energies and partial charges, and then training again with long-range energy contributions activated. Another application is transfer learning where we would for example pre-train a model on a large DFT dataset and than transfer the model to CCSD(T) accuracy on a smaller dataset. In order to facilitate the reproducibility of such multi-stage training, it can be fully defined in a single input file to the FeNNol training system. The following listing shows an example of a possible transfer-learning task:
\begin{verbatim}
training:
  batch_size: 128
  default_loss_type: mse
  lr: 1.e-3
  
  stages:
    train_dft:
      dspath: dft_dataset.pkl
      schedule_type: reduce_on_plateau
      end_event: [lr, 1.e-6]
      max_epochs: 100000
      loss:
        e:                        
          key: total_energy     
          ref: formation_energy   
          
    train_ccsdt:
       dspath: ccsdt_dataset.pkl
       schedule_type: cosine_onecycle
       end_event: none
       max_epochs: 100
       frozen: [embedding]
\end{verbatim}
In this example, the first stage \verb|train_dft| trains the model on formation energies from the (hypothetical) \verb|dft_dataset.pkl| dataset. This stage lasts a maximum of 100000 epochs or ends when the learning rate drops under $10^{-6}$. The second stage trains the model with a fixed schedule on the \verb|ccsdt_dataset.pkl|. Training parameters between stages are cumulative so that the loss did not need to be redefined in the second stage (provided that the keys in both datasets are the same). Finally, the second stage freezes the parameters of the model's embedding (defined by the module name \verb|embedding|) to limit the number of parameters to update for the transfer-learning task.\\

\paragraph*{Structure of accepted datasets} In its current version, FeNNol only accepts a simple but specific structure for datasets. The root of the dataset should be a dictionary with at least the two entries \verb|training| and \verb|validation| that are both lists containing the molecular configurations (one per element in the list) that will be used to train and validate the model respectively. Each molecular configuration is itself a dictionary which should contain the entries \verb|species| (an array of atomic numbers of the atoms in the configuration) and \verb|coordinates| (the array of 3D coordinates of the atoms), as well as all the keys used as reference in the definition of the training loss.
The dataset must be formatted by the user and can be provided in multiple formats. The simplest is to use the generic \verb|pickle| format to save the dataset from a python script. This format will be completely loaded in memory before training and loading batches is thus very fast. This format is best suited for small to moderately-large datasets as it must fit in memory and initial loading times can be very long for large datasets. In most applications, it should however be the simplest format and we found it practical even for quite large datasets (for example the ANI-2x dataset~\cite{devereux2020extending} with almost ten million entries).
For larger datasets, one can provide a hdf5 file or a sqlite database that can be queried without loading it in memory completely. These formats may however have slower access times.

\section{Running Molecular Dynamics simulations}\label{section:md}
FeNNol offers multiple ways of running molecular dynamics (MD) simulations with FeNNix models. One can either use custom python scripts, the provided calculator for the Atomic Simulation Environment~\cite{larsen2017atomic} (ASE) package, the Tinker-HP MD engine ~\cite{lagardere2018tinker,adjoua2021tinker} via the Deep-HP interface~\cite{inizan2022scalable} or FeNNol's native MD engine.

\subsection{Custom scripts}
FeNNix models define easy-to-use shorthand functions for computing energies, forces and stress tensor that can be used to build a custom MD engine in python. These can be accessed via the class methods: \verb|FENNIX.total_energy|, \verb|FENNIX.energy_and_forces| and \verb|FENNIX.energy_and_forces_and_virial|. They all accept keyword arguments which constitute the initial system state (so at least atomic species and coordinates) and return the total energy, forces and virial along with the final system state. Under the hood, these methods handle padding (and unpadding) of arrays to prevent too frequent recompilations of the model. Futhermore, these routines can be configured with the method \verb|FENNIX.set_energy_terms| to sum different keys of the final system state that are interpreted as energy components. For example, one could call 
\begin{verbatim}
  model.set_energy_terms(["nn_energy",
                          "coulomb_energy"])
\end{verbatim}
to define the total energy as the sum of a neural network potential (with \verb|nn_energy| the key referring to the output of the neural network) and electrostatic interactions via a Coulomb term (stored in the \verb|coulomb_energy| key).

\subsection{ASE calculator}
FeNNol provides a \verb|FENNIXCalculator| class that can be used with the Atomic Simulation Environment~\cite{larsen2017atomic} (ASE) package. This enables the use of all the advanced functionalities provided by ASE such as geometry optimization, phonon calculations and various MD algorithms in a very user-friendly manner. For example, a geometry optimization followed by a short MD simulation with a FeNNix model can be obtained with the following simple code:
\begin{verbatim}
    import ase
    from fennol.ase import FENNIXCalculator

    # initialization
    atoms = ase.io.read("aspirin.xyz")
    atoms.calc = FENNIXCalculator(
      "model_file.fnx"
    )

    # geometry optimization
    opt=ase.optimize.BFGS(atoms)
    opt.run(fmax=0.01)
    print(atoms.get_potential_energy())

    # MD simulation
    dyn = ase.md.verlet.VelocityVerlet(
            atoms, 0.5 * ase.units.fs)
    dyn.run(1000)
    
\end{verbatim}
Since ASE is able to communicate with the i-PI software~\cite{kapil2019pi}, simulations including nuclear quantum effects (NQEs) are directly available with this calculator.

\subsection{Deep-HP interface}
We modified the Deep-HP interface introduced in ref.~\onlinecite{inizan2022scalable} to enable the use of FeNNix models as energy components in Tinker-HP. This interface has multiple benefits. First, it allows the use of the different integrators, thermostats, barostats and accelerated sampling methods provided by Tinker-HP~\cite{lagardere2018tinker,adjoua2021tinker}. Efficient methods for simulating NQEs are also readily available with the Quantum-HP module~\cite{ple2022routine}. Second, it enables the parallelization of FeNNix models over multiple GPUs and multiple nodes, opening up simulations of very large systems with ML models. Note that currently, multi-GPU parallelization is only safe with purely local models such as ANI or Allegro. Finally, FeNNix models can be combined with traditional force fields, either via an ML/MM embedding~\cite{inizan2022scalable,lahey2020simulating,galvelis2023nnp}, or as energy components in a hybrid model~\cite{illarionov2023combining,wang2024incorporating}.

As in ref.~\onlinecite{inizan2022scalable}, we directly convert GPU pointers allocated from Tinker-HP (for example coordinates or neighbour lists) to Jax arrays via the DLPack protocol in order to minimize data movements and CPU/GPU communications. Note that the Tinker-HP/FeNNol interface is currently in early stages of development and will be refined and optimized in future works.

\subsection{Native MD engine}
FeNNol provides an efficient native engine to perform molecular dynamics simulations, which can be invoked with the \verb|fennol_md| command. The engine is written purely in Jax, is completely GPU-resident and leverages Jax's asynchronous dispatch capabilities for efficient dynamics even for small systems. Multiple thermostats are available (Langevin, Bussi-Parinello or Nose-Hoover) and the integration scheme used is based on symplectic time-splitting methods and is adapted depending on the thermostat (for example it reverts to the velocity Verlet integrator for NVE MD and to the BAOAB integrator~\cite{leimkuhler2013rational} for Langevin MD). Simulations can be performed either in vacuum or with periodic boundary conditions (with interactions with all images or minimum image convention). A current limitation of the engine, however, is that constant pressure simulations are not yet implemented. Furthermore, multi-GPU parallelization is not yet available.

For simulations including NQEs, we implemented two state-of-the-art methods: the adaptive Quantum Thermal Bath~\cite{mangaud2019fluctuation} (adQTB) and Ring-Polymer Molecular Dynamics~\cite{habershon2013ring} (RPMD). The adQTB method includes approximate NQEs in a classical-like simulations via a colored-noise Langevin thermostat. Our implementation incurs only a small overhead compared to classical Langevin MD (between 5\% and 20\% depending on system size, in agreement with previous studies~\cite{mauger2021nuclear}). On the other hand, RPMD uses replicas of the original system to achieve asymptotically exact (in the limit of infinite number of replicas) static properties of the quantum nuclei. This method is thus much more compute-intensive as it requires simulating a much larger system than in classical or adQTB MD (typical numbers of replicas are between 8 and 100 depending on the system and the simulated conditions). In our implementation, the potential energy of each of the replicas is efficiently obtained using FeNNol's batching capabilities.\\

\subsection{Performance benchmarks}\label{sec:perf}
In the remaining of this section, we provide a brief study of simulation accuracy with respect to Jax matrix multiplication level of precision (via energy conservation tests) and performance benchmarks for our implementation of the ANI-2x~\cite{devereux2020extending} model. The model is implemented only with standard FeNNol modules (\textit{i.e.} without custom CUDA kernels). Parameters for the neural network are obtained from the TorchANI~\cite{gao2020torchani} implementation and converted to Flax variables. We only use the first from the ensemble of eight neural networks that compose the ANI-2x model. All our tests were performed with Jax 0.4.25 with jaxlib 0.4.23 and CUDA 11.8. We used either a NVIDIA RTX 3090 or A100 GPU with respectively 24GB and 40GB of memory.

\begin{figure}
    \centering
    \includegraphics[width=0.5\textwidth]{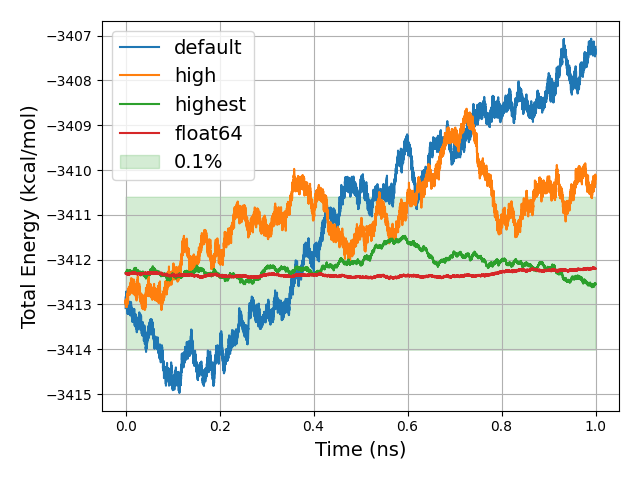}
    \caption{Total energy of a box of 216 water molecules as a function of time for different simulation precisions. All simulations use a timestep of 0.1fs and the ANI-2x model. The height of the green box represents 0.1\% of the initial total energy.}
    \label{fig:energy_conservation}
\end{figure}

For our accuracy tests, we compare single (\verb|float32|) and double precision (\verb|float64|) representations. For single precision, we also compare the various matrix multiplications precisions provided by Jax that are denoted as "default" (which uses \verb|bfloat16| numbers), "high" (which uses \verb|tensorfloat32|) and "highest" (which uses full \verb|float32|). Figure~\ref{fig:energy_conservation} shows the total energy of a periodic box of 216 water molecules over a 1~ns NVE simulation with a timestep of 0.1 fs (this small timestep was chosen to minimize integration error and focus on precision errors) with the different precision configurations. All the simulations start from the same off-equilibrium configuration and velocities are initialized to zero (this results in an average temperature of 315K with a standard deviation of 12K). It is clear from the figure, that for this simulation setup, both "default" and "high" configurations are not precise enough to obtain good energy conservation. Furthermore, we see a non-negligible shift in the initial energy compared to \verb|float64|. The "highest" \verb|float32| configuration provides acceptable energy conservation (fluctuations below 0.1\% of the initial energy, as indicated by the green box in the figure) and no initial energy shift compared to the full \verb|float64| simulation. From a performance perspective, differences between the three single precision configurations are small, while double precision is significantly more expensive on the RTX 3090 GPU (about a factor of 4). Note that this difference between single and double precision is less important with HPC GPUs. For example using double precision on a A100 GPU only adds a 15\% overhead. Given these tests, we set the default precision for FeNNol's MD engine to "highest" \verb|float32| and warn users of Jax that the default precision settings might not be sufficient in the context of MD simulations.\\

\begin{figure}
    \centering
    \includegraphics[width=0.48\textwidth]{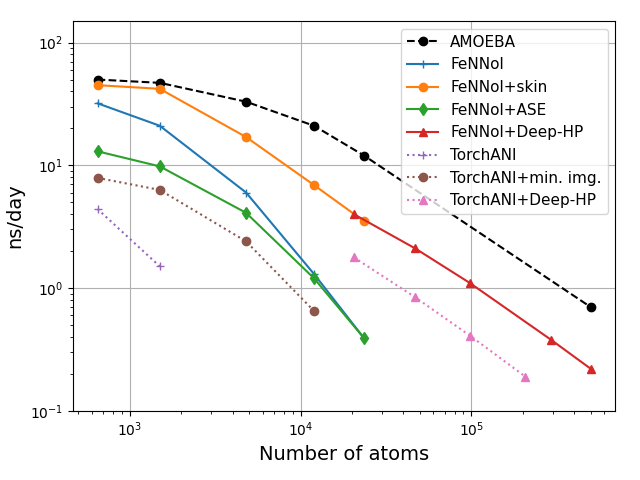}
    \caption{Performance (in ns simulated per day) for various system sizes. The black dashed line corresponds to the AMOEBA force field simulated using Tinker-HP. The solid lines correspond to ANI-2x with FeNNol's implementation and different MD engines. The dotted lines correspond to ANI-2x with TorchANI's implementation (min. img.=minimum image convention). All simulations use float32 precision, a 0.5fs timestep and a Langevin thermostat set at 300K. Simulations with TorchANI and FeNNol's native MD engine were performed on a RTX 3090 GPU. Simulations with Deep-HP were performed on a A100 GPU.}
    \label{fig:perf_vs_size}
\end{figure}

\begin{table*}
    \begin{tabular}{r |c|c|c|c|c|c}
     & FeNNol & FeNNol+skin & FeNNol+ASE & FeNNol+Deep-HP & TorchANI & TorchANI+Deep-HP \\
     \hline
    model implementation & Jax & Jax & Jax &Jax & Pytorch & Pytorch \\
    neighbour list & $\order{N^2}$ & $\order{N^2}/N_\text{update}$ & $\order{N^2}$ & $\order{N}$ & $ \order{N^2}$ & $\order{N}$ \\
    MD Engine & FeNNol(Jax) & FeNNol(Jax) & ASE(Numpy)  & Tinker-HP(CUDA)  & ASE(Numpy)& Tinker-HP(CUDA) \\
    \end{tabular}
    \caption{Summary of the different implementations that are compared in the performance benchmark of section~\ref{sec:perf}. The "neighbour list" line shows the complexity of the neighbour list construction. The column headers refer to the legend in Figure~\ref{fig:perf_vs_size}.}
    \label{table:implementation}
\end{table*}

We now test the performance of our implementation on water boxes of varying sizes (from 648 atoms to 500K atoms) and on the solvated DHFR (around 20K atoms). We compare our implementation (with ASE, Deep-HP or our native MD engine) to the base TorchANI implementation (with ASE or Deep-HP as MD engine). We also compare the performance of ANI-2x to the AMOEBA polarizable force field~\cite{ponder2010current} simulated using Tinker-HP. Table~\ref{table:implementation} summarizes the differences between the implementations that we compare in this section. All simulations use \verb|float32| precision, a 0.5 fs timestep, a Langevin thermostat set at 300K and periodic boundary conditions. Both FeNNol and TorchANI are set to only use the first out of the eight neural networks that compose the ANI-2x model. Simulations with TorchANI and FeNNol's native MD engine were performed on a RTX 3090 GPU with 24GB of memory. Simulations with FeNNol+Deep-HP were performed on a A100 GPU with 40GB of memory.

Figure~\ref{fig:perf_vs_size} shows the simulation performance (in ns simulated per day) for the different setups. In order to obtain fairer comparisons, we implemented the minimum image convention in TorchANI, which already provides a significant increase in performance for these systems. Note that we could not simulate the larger systems with TorchANI due to allocations exceeding our GPU's memory. If we first compare TorchANI to FeNNol both running with the ASE MD engine, we see that FeNNol gains almost a factor of two in performance which is mainly due to the efficient JIT compilation of Jax (note that a more direct comparison could be achieved by using the recently introduced JIT compilation tools in Pytorch 2 but that would require a complete rewrite of TorchANI with fixed array shapes). Since ASE integrates the equations of motion on the CPU, using this engine requires frequent data transfers between CPU and GPU. Using FeNNol's native MD engine, we alleviate this issue with a fully GPU-resident and asynchronous simulation, which considerably increase performance for smaller systems (nearly a factor of three for the smallest system). For larger systems, the main bottleneck becomes the construction of neighbour lists, which in the current implementation, scales quadratically with system size (compared to the linear scaling of the ANI-2x model itself). The impact of this quadratic scaling can be pushed back to larger system sizes by using a neighbour list "skin" 
and to reconstruct the full neighbour list only once every few steps. If we add a buffer of 2 \AA\ and, following the rule used in Tinker-HP, reconstruct the neighbour list once every 40 fs (80 steps), we obtain a significant improvement, reaching performance close to the AMOEBA force field on the smaller systems. This performance increase is due to both a gain in neighbour list construction and a better leveraging of Jax's asynchronous dispatch when using the additional buffer. 
Note that this comparison only takes into account the raw performance of the models with a fixed integration scheme. Typical MD runs with force fields however can usually afford larger timesteps and use advanced integration schemes (for example multi-timestepping\cite{lagardere2019pushing}) which are not compatible with current ML potentials and will require architectural evolutions. 

For systems of more than 20K atoms, the quadratic scaling of neighbour lists construction in FeNNol's native MD engine takes a significant toll on performance. We thus switched to Deep-HP -- which leverages the efficient cell lists implemented in Tinker-HP -- for systems between 20K and 500K atoms. For these large systems, we see similar scaling trends with FeNNol+Deep-HP compared to AMOEBA with about a factor of 3 in raw performance. Note that the FeNNol+Deep-HP interface is in very early development, so that we expect this gap to shrink with further optimizations and a tighter integration between both codes. In particular, we expect that performing all the model preprocessing (which does not require automatic differentiation) inside Tinker-HP will provide a boost in performance and significant memory savings. When comparing to the raw performance of TorchANI+Deep-HP, we see again a factor $\sim$2.5 with respect to FeNNol.

The results reported here were measured with a RTX 3090 GPU for smaller systems and a A100 HPC GPU for larger systems. We also tested the A100 GPU on the smaller systems and did not see significant differences with respect to the RTX GPU. For both GPUs, utilization is constantly close to 100\% for all systems except the smallest one. For the A100 GPU, power usage reaches 70\% of the cap value (280W / 400W) for the larger systems. For the RTX 3090, power usage stays close to the cap value (350W) for all systems except the smallest one, showing nearly maximal GPU occupancy. We can thus expect performance gains with the newer generation RTX 4090 that are close to theoretical values (i.e. at least a factor of two).

This brief study shows that neural network potentials can indeed reach performance close to force fields (at least the more expensive and accurate polarizable ones) and that the efficiency of the model implementation and the  MD engine are both critical. 



\section{Usage Examples}\label{section:examples}
In this section, we show three examples of models that can be trained with the FeNNol library and display training performance and timings on standard datasets. The first model is a message-passing CRATE model trained on the QM7-X dataset~\cite{hoja2021qm7}. The second model is an equivariant force-field-enhanced model similar to the FENNIX-OP1 model from ref.~\onlinecite{ple2023fennix} which is trained on the ANI-1x dataset~\cite{smith2020ani}. Finally, we show how FeNNol can be used to model other properties than the potential energy with the task of learning distributed multipoles from a subset of the SPICE datatset~\cite{eastman2023spice}. Note that the goal of this section is not to release new models but to give the reader an idea of the diversity of models that can be built with FeNNol, as well as the time and computer resources required for training.

\subsection{Training a CRATE model on the QM7-X dataset}\label{section:crate_qm7x}
For our first example, we train a CRATE model to reproduce formation energies and forces from the QM7-X dataset~\cite{hoja2021qm7}. The QM7-X dataset is composed of 4.2 million out-of-equilibrium configurations of small molecules (up to 7 heavy atoms with species H, C, N, O, S, Cl) with energies and forces computed at the PBE0+MDB level of theory. 
We trained two CRATE models with radial and angular resources, two or three layers of message-passing interactions with a cutoff radius of 5.3 \AA\ and short-range trainable ZBL repulsion~\cite{ziegler1985stopping}. Since this dataset only comprises isolated molecules, we did not include explicit long-range force-field terms in the model. The model was trained using batches of 128 configurations and a mean square error loss with a weight ratio of 1:100 between total energies and  forces. We refer the reader to the archive file provided in Supplementary Material for the complete input file defining the model, training parameters and loss function.
We used the Adabelief~\cite{zhuang2020adabelief} optimizer with a fixed learning rate schedule starting form 2.e-4 and decreasing to 1.e-6 over the course of $10^6$ random batches. We reserved a random 20\% of the dataset for validation. Training completed in 3.7 hours and 4 hours for the two- and three-layers model respectively, on a single NVIDIA RTX 3090 GPU. The two-layers model achieved a mean absolute error (MAE) on the validation set of 0.70 kcal/mol/\AA\ for forces and 0.59 kcal/mol on energies. The three-layer model achieved a slightly better MAE of 0.49 kcal/mol/\AA\ on forces and 0.39 kcal/mol on energies. These results are in line with those previously reported in ref.~\onlinecite{unke2021spookynet} for the PaiNN and SpookyNet models and show that state-of-the-art results can be achieved in a few hours of training on a single commodity GPU.

\subsection{Force-field-enhanced equivariant model}
For this second example, we trained a model similar to the FENNIX-OP1 model in ref.~\cite{ple2023fennix}. We use an Allegro embedding~\cite{musaelian2022learning} with 256 scalar features, 16 equivariant channels with maximum degree $l_\text{max}=2$ and three local interaction layers with a cutoff of 5.3\AA. For this model, we included four energy components: local pairwise energies obtained via a neural network, a short-range ZBL repulsion energy, Coulomb interactions using fluctuating atom-centered charges and dispersion interactions via the two-body Optimized Quantum Drude Oscillator model~\cite{khabibrakhmanov2023universal}. The Coulomb and dispersion interactions are parameterized via charges and atomic volumes learned from MB-ISA~\cite{verstraelen2016minimal} density partitioning provided in the dataset. The sum of partial charges is constrained to the total charge of the system using a neural charge equilibration inspired from ref.~\onlinecite{zubatyuk2021teaching}. We train the model on the ANI-1x dataset~\cite{smith2020ani} using the same training setup as in section~\ref{section:crate_qm7x}. The weight ratio on the different loss components is 1:100:5:5 for total energies, force components, charges and volume ratios respectively. Training completed in 18 hours with a MAE of 1.10 kcal/mol/\AA on forces, 0.85 kcal/mol on total energies, 0.0060 $e$ on charges and 0.0059 on volume ratios. 

For comparison, we trained a CRATE model with radial and equivariant resources and two message-passing layers with a cutoff of 5.3\AA\ (yielding a receptive field of 10.6\AA). Similarly to the Allegro model, we used 16 equivariant channels with $l_\text{max}=2$. At each layer, we perform two atom-centered channel-wise tensor products. Training completed in 5 hours which is comparable to the models in the previous section and much faster than for the Allegro model. This difference is mainly due to the pairwise nature of the Allegro model which is more computationally demanding than an atom-centered model. The CRATE model achieved an MAE of 1.14 kcal/mol/\AA\ on forces, 0.81 kcal/mol on total energies, 0.0050 $e$ on charges and 0.0043 on volume ratios, showing similar performance to the Allegro model.

\subsection{Learning distributed multipoles}
For the last example, we trained an equivariant model to reproduce distributed MB-ISA multipoles (up to quadrupoles) from the Pubchem subset of the SPICE dataset~\cite{eastman2023spice}. We selected a CRATE embedding with radial, angular and equivariant resources with three interaction layers of cutoff 3.5\AA. We used 32 equivariant channels with maximum degree $l_\text{max}=2$ and perform one tensor product per layer. We obtain the atomic dipoles and quadrupoles from a linear combination of the equivariant channels of the corresponding irreducible representation. To obtain total-charge-constrained partial charges, we use the scalar embedding produced by CRATE to define a weight on each atom that is then used to distribute all the valence electrons in the system (including eventual ionic charges). This constrains the total number of electrons and prevents the atoms from unphysically depleting more than their number of valence electrons with too large positive charges. We kept 10\% of the 1.4 million configurations as a validation set. Furthermore, we preprocessed the quadrupoles in the dataset to make them traceless. We used a similar training setup as in the previous two examples and applied the same weight to all the multipole components in the loss function. Training completed in 14 hours and the model achieved a MAE of 0.004 $e$ on charges, 0.0017 $e$\AA\ on dipole components and 0.0013 $e$\AA$^2$ on quadrupole components on the validation set. For reference, the MAE for a trivial model that predicts only neutral spherical atoms is 0.27 $e$ for charges, 0.019 $e$\AA\  for dipoles and $0.0087$ $e$\AA$^2$ for quadrupoles.  

\section{Conclusion}
We introduced FeNNol, a new library for building, training and running molecular dynamics simulations with machine-learning potentials, with a particular focus on force-field-enhanced neural network potentials. It leverages the Jax library~\cite{jax2018github} and the Flax~\cite{flax2020github} machine-learning framework to construct parameterized functions that efficiently run on accelerators and can be automatically differentiated. FeNNol enables the user to build complex models in a declarative way, without the need for explicit programming. The library provides an extensive collection of modules that cover atomic neighborhood descriptors (or embeddings), various architectures of neural networks, encoding schemes for spatial and chemical information, operation modules and physics modules (for example force-field interaction terms). These modules can be composed to build models that are tailored to the user's requirements. More advanced users can also register custom modules that can be made very efficient without resorting to low-level implementation thanks to Jax's JIT compilation capabilities.

In order to showcase the highly configurable nature of FeNNol modules, we developed the CRATE atomic embedding that can be set up to exploit information from various chemical and geometric resources (radial basis, angular information, E(3) equivariant tensors, etc...) and combine them via either local or message-passing interaction layers. This embedding can thus be adjusted by the user depending on the complexity of the learning task and the computational cost requirements.

FeNNol provides a complete training system that allows defining multi-objective and multi-stage training in an easy and reproducible way. We showed with a few examples that it can be used to train models for different tasks (potential energy surfaces, atom-in-molecule properties, distributed multipoles) with state-of-the-art results in a few hours on a single commodity GPU.

Finally, we provide three main approaches for running MD simulations with ML models: an ASE calculator, an interface to Tinker-HP via Deep-HP~\cite{inizan2022scalable} and an efficient native MD engine. The ASE calculator allows to transparently use all of the functionalities and benefit from the flexibility of the Atomic Simulation Environment~\cite{larsen2017atomic}. Similarly, the Deep-HP interface enables the use of all of Tinker-HP's advanced functionalities and its massive parallelization capabilities. Furthermore it allows for mixed ML/MM simulations as shown in ref.~\onlinecite{inizan2022scalable}. We showed that our early verision of the interface enables simulations of systems up to 500K atoms on a single A100 GPU with performance and scaling comparable to the AMOEBA force fields. Finally, FeNNol's native MD engine allows to perform very efficient MD simulations of intermediate size systems (more than 20K atoms) directly within the FeNNol package by leveraging the full capabilities of Jax's JIT compilation and asynchronous dispatch. We showed that our implementation of the ANI-2x model can reach simulation speeds nearly on par with an optimized version of the AMOEBA polarizable force field.

FeNNol thus constitutes a very modular framework for modeling complex chemical systems with physics-enhanced machine-learning models, allowing to easily experiment with different architectures and to iterate quickly, while remaining highly efficient. We hope that this library will accelerate the development and application of ML models and facilitate their access to a wider audience.

\section{Code Availability}
The FeNNol library is available on Github \url{https://github.com/thomasple/FeNNol} under the LGPL3 licence. Example input files are provided both for model training and molecular dynamics simulations.

\section{Data Availability}
The input and output files for the three examples of section~\ref{section:examples} are provided in an archive file. Datasets used in these examples are publicly available and can be freely obtained from the relevant repositories. The preprocessed dataset files (formatted and with the specific data split used here) can be provided upon request.

\section{Acknowledgement}
This work was made possible thanks to funding from the European Research Council (ERC) under the European Union's Horizon 2020 research and innovation program (grant agreement No 810367), project EMC2. Computations have been performed at GENCI (IDRIS, Orsay, France and TGCC, Bruyères le Chatel) on grant no A0130712052. 

\appendix
\section{Detailed description of the CRATE embedding}\label{appendix:crate}
CRATE defines an embeddding vector for each atom in the system that is iteratively refined over $L$ layers by combining chemical and geometric "resources" from its local environment. The embedding $x_i$ for atom $i$ is initialized with an encoding of its chemical species $e_{Zi}$ (which can be any of the species encodings provided in FeNNol).

At each interaction layer $l$, the embedding is projected to two lower-dimensional spaces $r_i$,$s_i$ via a learnable affine transform:
\begin{align}
    r_i &= W_r^{(l)} x_i + b_r^{(l)}\\
    s_i &= W_s^{(l)} x_i + b_s^{(l)}
\end{align}
The $r_i$ vector is used to retain information from the previous layer while the $s_i$ vector will be combined with local resources. For each neighbour $j$ of atom $i$, we then form a neighbour embedding $s_{ij}$. When using message passing, this neighbour embedding is simply $s_j$. Alternatively, when a fully local model is required, we need to form a neighbour embedding that accounts both for the current embedding of atom $i$ and available local information on atom $j$, i.e. its chemical species. For efficiency's sake, we use a key-query mechanism similar to the one used in attention:
\begin{align}
    Q_i &= W_Q^{(l)} s_i \quad\text{;}\quad K_j = W_K^{(l)} e_{Z_j}\\
    s_{ij} &= Q_i^T K_j / \sqrt{d}
\end{align}
with $W_Q^{(l)} \in \mathbb{R}^{d\times \text{dim}(s)}$ and $W_K^{(l)} \in \mathbb{R}^{d\times \text{dim}(e_Z)}$ two trainable matrices (with $d$ an hyperparameter intermediate dimension). 

The local geometric resources are then computed and concatenated with $r_i$ to form a vector $R_i$ containing all the resources from this layer. 
\begin{equation}\label{eq:resource_concat}
    R_i = r_i~||~\Big(\underset{g\in\text{res(l)}}{||} g(x_i,\{s_{ij}\})\Big)
\end{equation}
These resources are then mixed via a multi-layer perceptron $MLP^{(l)}$ to form the embedding update:
\begin{equation}
    x_i^{(l+1)} = \sigma(F^{(l)}) \odot x_i^{(l)} + u\qty(MLP^{(l)}(R_i))
\end{equation}
where $u$ is an optional activation function, $F^{(l)} \in \mathbb{R}^{\text{dim}(x)}$ and $\sigma(F^{(l)})$  is a trainable "forget gate" (with $\sigma$ the sigmoid function and $\odot$ the element-wise multiplication).
The current implementation has four available resources: radial, angular, equivariant and long-range.\\

\paragraph*{Radial resources} radial resources are simply obtained by an outer product of the neighbour embedding with a projection of the distance into a radial basis $B(r_{ij})$. These pairwise chemical-radial information are then multiplied by a switching function $f(r_{ij})$ to ensure continuity when atoms enter or leave the neighborhood and then summed over the neighbours:
\begin{equation}
    g^\ttiny{rad}_i = \sum_j s_{ij}\otimes B(r_{ij}) f(r_{ij})
\end{equation}
The radial basis and the switching function can be chosen by the user from FeNNol's provided functional forms.

\paragraph*{Angular resources} angular resources are obtained by combining information from triplets of atoms. Importantly, the user can choose the cutoff defining the neighborhood for triplets of atoms to be smaller than for the radial resources. This is significant in the condensed phase as the number of triplets of atoms quickly grows with the cutoff radius. To build angular resources, we start by forming a reduced chemical-radial basis $D_{ij}$ and $D_{ik}$ for the two edges:
\begin{equation}
   \big[D_{ij}\big]_c = \tilde{f}(r_{ij}) \sum_{ab}  \big[\tilde{B}(r_{ij})\big]_a~\big[s_{ij}\big]_b~\big[W^{(l)}_\ttiny{ang}\big]_{abc}
\end{equation}
with $W^{(l)}_\ttiny{ang} \in \mathbb{R}^{\text{dim}(\tilde{B})\times \text{dim}(s)\times d_a}$ a trainable tensor and $d_a$ a hyperparameter of the model. $\tilde{f}$ is the switching function adapted to the cutoff defined for triplets and $\tilde{B}$ is a radial basis for the triplet graph (that can be identical to $B$ or not). We then project the angle between the two edges into an angular basis $\Theta_{ijk}$ (this basis can be either a small fourier expansion $[\Theta_{ijk}]_n=\cos(n\theta_{ijk})$ or the angular symmetry function used in the ANI models~\cite{smith2017ani} $[\Theta_{ijk}]_n=(0.5+0.5\cos(\theta_{ijk}-\theta_n))^\zeta$) and combine the information as:
\begin{equation}
    g_i^\ttiny{ang} = \sum_{\{jk\}} \Theta_{ijk} \otimes \qty(D_{ij} \odot D_{ik})
\end{equation}

\paragraph*{equivariant resources} equivariant resources are obtained by forming an equivariant neighborhood basis from a chemical-radial basis and spherical harmonics and iteratively refining an equivariant embedding via multiple atom-centerd tensor products, similarly to what is proposed in the MACE model~\cite{batatia2022mace}. The equivariant neighborhood basis $\hat{\rho}_i$ (we denote equivariant tensors with hat notation) is initialized to zero and updated at each layer with:
\begin{align}
    b_{ij}^{(l)} &= W_e^{(l)}\Big(s_{ij} \otimes B(r_{ij})f(r_{ij})\Big)\\
    \hat{V}_{ij}^{(l)} &= \left\{\begin{matrix}
        \hat{V}_j^{(l-1)}  &\text{if } l>1 \text{ and message-passing,}\\
        \text{else } \hat{Y}_{ij}
    \end{matrix}\right.\\
    \hat{\rho}_i^{(l)} &=  \hat{\rho}_i^{(l-1)} + \sum_j b_{ij}^{(l)} \otimes \hat{V}_{ij}^{(l)}
\end{align}
with $\hat{V}_j$ the current equivariant embedding of atom $j$ and $\hat{Y}_{ij}$ the tensor representing the coefficients of the direction between atom $i$ and $j$ in the real spherical harmonics basis up to a user-defined maximum degree $l_\text{max}$. The trainable matrix $ W_e^{(l)}$ converts the radial-chemical basis to a user-selected $N_{channels}$ equivariant channels. This number of channels is typically kept small to limit the high computational cost of equivariant tensor products. For the first interaction layer, we initialize the equivariant embedding $\hat{V}_i$ as a linear combination of the irreps of each channel: $[V_i^{(0)}]_{n\lambda m}~=~\sum_{n'}~[W_v^{(0)}]_{nn'\lambda} [Y_{ij}]_{n'\lambda m}$.

After forming the neighborhood basis, we perform a series of $N_\ttiny{TP}$ tensor products to update $V_i$:
\begin{widetext}
\begin{align}
    \qty[\hat{L}_i]_{n\lambda m}^{(l,t)} &= \sum_{p=(\lambda_1,\lambda_2\rightarrow \lambda)} \qty[W_\ttiny{path}^{(l,t)}]_p\qty(\sum_{m_1,m_2}C^{\lambda_1\lambda_2\lambda}_{m_1 m_2 m}~\qty[\hat{\rho}_i^{(l)}]_{n\lambda_1 m_1}~[\hat{V}_i^{(l,t-1)}]_{n\lambda_2 m_2})\\
    \qty[\hat{V}_i^{(l,t)}]_{n\lambda m} &= \sum_{n'} \qty[W_v^{(l,t)}]_{nn'} \qty[\hat{V}_i^{(l,t-1)} + \hat{L}_i^{(l,t)}]_{n'\lambda m}\label{eq:update_Vi}
\end{align}
\end{widetext}
where $C^{\lambda_1\lambda_2\lambda}_{m_1 m_2 m}$ are Clebsch–Gordan coefficients for the basis of real spherical harmonics, the index $p$ enumerates tensor product paths that produce a certain irrep of degree $\lambda$ (the user can choose to include either all the paths that produce a certain $\lambda$ or only the ones which conserve the parity of the output irrep). The trainable parameters $W_\ttiny{path}^{(l,t)}$ combine the different tensor product paths to conserve the number of irreps. This is similar to the scheme proposed in ref.~\onlinecite{kozinsky2023scaling} and is key to prevent the explosion of the number of paths after iterating tensor products. Equation~\eqref{eq:update_Vi} corresponds to an update of $V_i$ followed by channel mixing. We initialize as $\hat{V}_i^{(l,0)}=\hat{V}_i^{(l-1)}$ and keep the last iteration as the current equivariant embedding $\hat{V}_i^{(l)}=\hat{V}_i^{(l,N_\ttiny{TP})}$.

Finally, equivariant resources at layer $l$ are obtained by concatenating the scalar irreps from the different $\hat{L}_i^{(l,t)}$ as:
\begin{equation}
    g_i^\text{E3} = \underset{t,n}{||} ~ [\hat{L}_i^{(l,t)}]_{n00}
\end{equation}

\paragraph*{Long-range resources}  long-range resources on atom $i$ are obtained as the electrostatic potential generated by a set of fictitious point charges placed on all the other atoms. We start by obtaining $N_{q}$ sets of point charges and short-range damping coefficients as:
\begin{align}
    \qty[q_i]_n,\qty[\tilde{\gamma}_i]_n  &= \qty[\text{MLP}^{(q)}(e_{Z_i})]_n\\
    \gamma_i &= \text{softplus}(\tilde{\gamma}_i)
\end{align}
where $\text{MLP}^{(q)}$ is a multi-layer perceptron outputting $2N_{q}$ numbers.
For each atom, we also have $N_{q}$ trainable effective radii $a_i$ that are initialized to the atom's Van der Waals radius and we obtain a pairwise damping as $\gamma_{ij}=(a_i^2 + a_j^2)^{-1/2}$. The long-range resources are then obtained as:
\begin{equation}
    \qty[g_i^{lr}]_n = \sum_j \qty[q_j]_n \frac{\text{erf}\Big(\qty[\gamma_i]_n\gamma_{ij} r_{ij}\Big)}{r_{ij}}
\end{equation}
This scheme is inspired from the LODE framework~\cite{grisafi2019incorporating,grisafi2021multi} but we simplified it to keep only invariant quantities instead of full equivariant tensors. 
In the current implementation, long-range resources are only computed at the first iteration and fed back unchanged into the embedding update of equation~\eqref{eq:resource_concat} at each layer. This avoids multiple iterations of costly long-range interactions. For now, the long-range embedding is not compatible with periodic boundary conditions.


\bibliography{biblio.bib}

\begin{thebibliography}{103}%
\makeatletter
\providecommand \@ifxundefined [1]{%
 \@ifx{#1\undefined}
}%
\providecommand \@ifnum [1]{%
 \ifnum #1\expandafter \@firstoftwo
 \else \expandafter \@secondoftwo
 \fi
}%
\providecommand \@ifx [1]{%
 \ifx #1\expandafter \@firstoftwo
 \else \expandafter \@secondoftwo
 \fi
}%
\providecommand \natexlab [1]{#1}%
\providecommand \enquote  [1]{``#1''}%
\providecommand \bibnamefont  [1]{#1}%
\providecommand \bibfnamefont [1]{#1}%
\providecommand \citenamefont [1]{#1}%
\providecommand \href@noop [0]{\@secondoftwo}%
\providecommand \href [0]{\begingroup \@sanitize@url \@href}%
\providecommand \@href[1]{\@@startlink{#1}\@@href}%
\providecommand \@@href[1]{\endgroup#1\@@endlink}%
\providecommand \@sanitize@url [0]{\catcode `\\12\catcode `\$12\catcode `\&12\catcode `\#12\catcode `\^12\catcode `\_12\catcode `\%12\relax}%
\providecommand \@@startlink[1]{}%
\providecommand \@@endlink[0]{}%
\providecommand \url  [0]{\begingroup\@sanitize@url \@url }%
\providecommand \@url [1]{\endgroup\@href {#1}{\urlprefix }}%
\providecommand \urlprefix  [0]{URL }%
\providecommand \Eprint [0]{\href }%
\providecommand \doibase [0]{http://dx.doi.org/}%
\providecommand \selectlanguage [0]{\@gobble}%
\providecommand \bibinfo  [0]{\@secondoftwo}%
\providecommand \bibfield  [0]{\@secondoftwo}%
\providecommand \translation [1]{[#1]}%
\providecommand \BibitemOpen [0]{}%
\providecommand \bibitemStop [0]{}%
\providecommand \bibitemNoStop [0]{.\EOS\space}%
\providecommand \EOS [0]{\spacefactor3000\relax}%
\providecommand \BibitemShut  [1]{\csname bibitem#1\endcsname}%
\let\auto@bib@innerbib\@empty
\bibitem [{\citenamefont {Wang}\ \emph {et~al.}(2004)\citenamefont {Wang}, \citenamefont {Wolf}, \citenamefont {Caldwell}, \citenamefont {Kollman},\ and\ \citenamefont {Case}}]{wang2004development}%
  \BibitemOpen
  \bibfield  {author} {\bibinfo {author} {\bibfnamefont {J.}~\bibnamefont {Wang}}, \bibinfo {author} {\bibfnamefont {R.~M.}\ \bibnamefont {Wolf}}, \bibinfo {author} {\bibfnamefont {J.~W.}\ \bibnamefont {Caldwell}}, \bibinfo {author} {\bibfnamefont {P.~A.}\ \bibnamefont {Kollman}}, \ and\ \bibinfo {author} {\bibfnamefont {D.~A.}\ \bibnamefont {Case}},\ }\href@noop {} {\bibfield  {journal} {\bibinfo  {journal} {Journal of computational chemistry}\ }\textbf {\bibinfo {volume} {25}},\ \bibinfo {pages} {1157} (\bibinfo {year} {2004})}\BibitemShut {NoStop}%
\bibitem [{\citenamefont {Vanommeslaeghe}\ \emph {et~al.}(2010)\citenamefont {Vanommeslaeghe}, \citenamefont {Hatcher}, \citenamefont {Acharya}, \citenamefont {Kundu}, \citenamefont {Zhong}, \citenamefont {Shim}, \citenamefont {Darian}, \citenamefont {Guvench}, \citenamefont {Lopes}, \citenamefont {Vorobyov} \emph {et~al.}}]{vanommeslaeghe2010charmm}%
  \BibitemOpen
  \bibfield  {author} {\bibinfo {author} {\bibfnamefont {K.}~\bibnamefont {Vanommeslaeghe}}, \bibinfo {author} {\bibfnamefont {E.}~\bibnamefont {Hatcher}}, \bibinfo {author} {\bibfnamefont {C.}~\bibnamefont {Acharya}}, \bibinfo {author} {\bibfnamefont {S.}~\bibnamefont {Kundu}}, \bibinfo {author} {\bibfnamefont {S.}~\bibnamefont {Zhong}}, \bibinfo {author} {\bibfnamefont {J.}~\bibnamefont {Shim}}, \bibinfo {author} {\bibfnamefont {E.}~\bibnamefont {Darian}}, \bibinfo {author} {\bibfnamefont {O.}~\bibnamefont {Guvench}}, \bibinfo {author} {\bibfnamefont {P.}~\bibnamefont {Lopes}}, \bibinfo {author} {\bibfnamefont {I.}~\bibnamefont {Vorobyov}},  \emph {et~al.},\ }\href@noop {} {\bibfield  {journal} {\bibinfo  {journal} {Journal of computational chemistry}\ }\textbf {\bibinfo {volume} {31}},\ \bibinfo {pages} {671} (\bibinfo {year} {2010})}\BibitemShut {NoStop}%
\bibitem [{\citenamefont {Jorgensen}, \citenamefont {Maxwell},\ and\ \citenamefont {Tirado-Rives}(1996)}]{OPLS}%
  \BibitemOpen
  \bibfield  {author} {\bibinfo {author} {\bibfnamefont {W.~L.}\ \bibnamefont {Jorgensen}}, \bibinfo {author} {\bibfnamefont {D.~S.}\ \bibnamefont {Maxwell}}, \ and\ \bibinfo {author} {\bibfnamefont {J.}~\bibnamefont {Tirado-Rives}},\ }\href {\doibase 10.1021/ja9621760} {\bibfield  {journal} {\bibinfo  {journal} {Journal of the American Chemical Society}\ }\textbf {\bibinfo {volume} {118}},\ \bibinfo {pages} {11225} (\bibinfo {year} {1996})},\ \Eprint {http://arxiv.org/abs/https://doi.org/10.1021/ja9621760} {https://doi.org/10.1021/ja9621760} \BibitemShut {NoStop}%
\bibitem [{\citenamefont {Gresh}\ \emph {et~al.}(2007)\citenamefont {Gresh}, \citenamefont {Cisneros}, \citenamefont {Darden},\ and\ \citenamefont {Piquemal}}]{gresh2007anisotropic}%
  \BibitemOpen
  \bibfield  {author} {\bibinfo {author} {\bibfnamefont {N.}~\bibnamefont {Gresh}}, \bibinfo {author} {\bibfnamefont {G.~A.}\ \bibnamefont {Cisneros}}, \bibinfo {author} {\bibfnamefont {T.~A.}\ \bibnamefont {Darden}}, \ and\ \bibinfo {author} {\bibfnamefont {J.-P.}\ \bibnamefont {Piquemal}},\ }\href@noop {} {\bibfield  {journal} {\bibinfo  {journal} {Journal of Chemical Theory and Computation}\ }\textbf {\bibinfo {volume} {3}},\ \bibinfo {pages} {1960} (\bibinfo {year} {2007})}\BibitemShut {NoStop}%
\bibitem [{\citenamefont {Naseem-Khan}\ \emph {et~al.}(2022)\citenamefont {Naseem-Khan}, \citenamefont {Lagard{\`e}re}, \citenamefont {Narth}, \citenamefont {Cisneros}, \citenamefont {Ren}, \citenamefont {Gresh},\ and\ \citenamefont {Piquemal}}]{naseem2022development}%
  \BibitemOpen
  \bibfield  {author} {\bibinfo {author} {\bibfnamefont {S.}~\bibnamefont {Naseem-Khan}}, \bibinfo {author} {\bibfnamefont {L.}~\bibnamefont {Lagard{\`e}re}}, \bibinfo {author} {\bibfnamefont {C.}~\bibnamefont {Narth}}, \bibinfo {author} {\bibfnamefont {G.~A.}\ \bibnamefont {Cisneros}}, \bibinfo {author} {\bibfnamefont {P.}~\bibnamefont {Ren}}, \bibinfo {author} {\bibfnamefont {N.}~\bibnamefont {Gresh}}, \ and\ \bibinfo {author} {\bibfnamefont {J.-P.}\ \bibnamefont {Piquemal}},\ }\href@noop {} {\bibfield  {journal} {\bibinfo  {journal} {arXiv preprint arXiv:2201.00804}\ } (\bibinfo {year} {2022})}\BibitemShut {NoStop}%
\bibitem [{\citenamefont {Piquemal}\ \emph {et~al.}(2006)\citenamefont {Piquemal}, \citenamefont {Cisneros}, \citenamefont {Reinhardt}, \citenamefont {Gresh},\ and\ \citenamefont {Darden}}]{piquemal2006towards}%
  \BibitemOpen
  \bibfield  {author} {\bibinfo {author} {\bibfnamefont {J.-P.}\ \bibnamefont {Piquemal}}, \bibinfo {author} {\bibfnamefont {G.~A.}\ \bibnamefont {Cisneros}}, \bibinfo {author} {\bibfnamefont {P.}~\bibnamefont {Reinhardt}}, \bibinfo {author} {\bibfnamefont {N.}~\bibnamefont {Gresh}}, \ and\ \bibinfo {author} {\bibfnamefont {T.~A.}\ \bibnamefont {Darden}},\ }\href@noop {} {\bibfield  {journal} {\bibinfo  {journal} {The Journal of chemical physics}\ }\textbf {\bibinfo {volume} {124}} (\bibinfo {year} {2006})}\BibitemShut {NoStop}%
\bibitem [{\citenamefont {Duke}\ \emph {et~al.}(2014)\citenamefont {Duke}, \citenamefont {Starovoytov}, \citenamefont {Piquemal},\ and\ \citenamefont {Cisneros}}]{duke2014gem}%
  \BibitemOpen
  \bibfield  {author} {\bibinfo {author} {\bibfnamefont {R.~E.}\ \bibnamefont {Duke}}, \bibinfo {author} {\bibfnamefont {O.~N.}\ \bibnamefont {Starovoytov}}, \bibinfo {author} {\bibfnamefont {J.-P.}\ \bibnamefont {Piquemal}}, \ and\ \bibinfo {author} {\bibfnamefont {G.~A.}\ \bibnamefont {Cisneros}},\ }\href@noop {} {\bibfield  {journal} {\bibinfo  {journal} {Journal of Chemical Theory and Computation}\ }\textbf {\bibinfo {volume} {10}},\ \bibinfo {pages} {1361} (\bibinfo {year} {2014})}\BibitemShut {NoStop}%
\bibitem [{\citenamefont {Reddy}\ \emph {et~al.}(2016)\citenamefont {Reddy}, \citenamefont {Straight}, \citenamefont {Bajaj}, \citenamefont {Huy~Pham}, \citenamefont {Riera}, \citenamefont {Moberg}, \citenamefont {Morales}, \citenamefont {Knight}, \citenamefont {G{\"o}tz},\ and\ \citenamefont {Paesani}}]{reddy2016accuracy}%
  \BibitemOpen
  \bibfield  {author} {\bibinfo {author} {\bibfnamefont {S.~K.}\ \bibnamefont {Reddy}}, \bibinfo {author} {\bibfnamefont {S.~C.}\ \bibnamefont {Straight}}, \bibinfo {author} {\bibfnamefont {P.}~\bibnamefont {Bajaj}}, \bibinfo {author} {\bibfnamefont {C.}~\bibnamefont {Huy~Pham}}, \bibinfo {author} {\bibfnamefont {M.}~\bibnamefont {Riera}}, \bibinfo {author} {\bibfnamefont {D.~R.}\ \bibnamefont {Moberg}}, \bibinfo {author} {\bibfnamefont {M.~A.}\ \bibnamefont {Morales}}, \bibinfo {author} {\bibfnamefont {C.}~\bibnamefont {Knight}}, \bibinfo {author} {\bibfnamefont {A.~W.}\ \bibnamefont {G{\"o}tz}}, \ and\ \bibinfo {author} {\bibfnamefont {F.}~\bibnamefont {Paesani}},\ }\href@noop {} {\bibfield  {journal} {\bibinfo  {journal} {The Journal of Chemical Physics}\ }\textbf {\bibinfo {volume} {145}},\ \bibinfo {pages} {194504} (\bibinfo {year} {2016})}\BibitemShut {NoStop}%
\bibitem [{\citenamefont {Zhu}\ \emph {et~al.}(2023)\citenamefont {Zhu}, \citenamefont {Riera}, \citenamefont {Bull-Vulpe},\ and\ \citenamefont {Paesani}}]{zhu2023mb}%
  \BibitemOpen
  \bibfield  {author} {\bibinfo {author} {\bibfnamefont {X.}~\bibnamefont {Zhu}}, \bibinfo {author} {\bibfnamefont {M.}~\bibnamefont {Riera}}, \bibinfo {author} {\bibfnamefont {E.~F.}\ \bibnamefont {Bull-Vulpe}}, \ and\ \bibinfo {author} {\bibfnamefont {F.}~\bibnamefont {Paesani}},\ }\href@noop {} {\bibfield  {journal} {\bibinfo  {journal} {Journal of Chemical Theory and Computation}\ } (\bibinfo {year} {2023})}\BibitemShut {NoStop}%
\bibitem [{\citenamefont {Ren}\ and\ \citenamefont {Ponder}(2003)}]{AMOEBA03}%
  \BibitemOpen
  \bibfield  {author} {\bibinfo {author} {\bibfnamefont {P.}~\bibnamefont {Ren}}\ and\ \bibinfo {author} {\bibfnamefont {J.~W.}\ \bibnamefont {Ponder}},\ }\href {\doibase 10.1021/jp027815+} {\bibfield  {journal} {\bibinfo  {journal} {The Journal of Physical Chemistry B}\ }\textbf {\bibinfo {volume} {107}},\ \bibinfo {pages} {5933} (\bibinfo {year} {2003})},\ \Eprint {http://arxiv.org/abs/https://doi.org/10.1021/jp027815+} {https://doi.org/10.1021/jp027815+} \BibitemShut {NoStop}%
\bibitem [{\citenamefont {Ponder}\ \emph {et~al.}(2010)\citenamefont {Ponder}, \citenamefont {Wu}, \citenamefont {Ren}, \citenamefont {Pande}, \citenamefont {Chodera}, \citenamefont {Schnieders}, \citenamefont {Haque}, \citenamefont {Mobley}, \citenamefont {Lambrecht}, \citenamefont {DiStasio~Jr} \emph {et~al.}}]{ponder2010current}%
  \BibitemOpen
  \bibfield  {author} {\bibinfo {author} {\bibfnamefont {J.~W.}\ \bibnamefont {Ponder}}, \bibinfo {author} {\bibfnamefont {C.}~\bibnamefont {Wu}}, \bibinfo {author} {\bibfnamefont {P.}~\bibnamefont {Ren}}, \bibinfo {author} {\bibfnamefont {V.~S.}\ \bibnamefont {Pande}}, \bibinfo {author} {\bibfnamefont {J.~D.}\ \bibnamefont {Chodera}}, \bibinfo {author} {\bibfnamefont {M.~J.}\ \bibnamefont {Schnieders}}, \bibinfo {author} {\bibfnamefont {I.}~\bibnamefont {Haque}}, \bibinfo {author} {\bibfnamefont {D.~L.}\ \bibnamefont {Mobley}}, \bibinfo {author} {\bibfnamefont {D.~S.}\ \bibnamefont {Lambrecht}}, \bibinfo {author} {\bibfnamefont {R.~A.}\ \bibnamefont {DiStasio~Jr}},  \emph {et~al.},\ }\href@noop {} {\bibfield  {journal} {\bibinfo  {journal} {The Journal of Physical Chemistry B}\ }\textbf {\bibinfo {volume} {114}},\ \bibinfo {pages} {2549} (\bibinfo {year} {2010})}\BibitemShut {NoStop}%
\bibitem [{\citenamefont {Liu}, \citenamefont {Piquemal},\ and\ \citenamefont {Ren}(2019)}]{AMOEBA+1}%
  \BibitemOpen
  \bibfield  {author} {\bibinfo {author} {\bibfnamefont {C.}~\bibnamefont {Liu}}, \bibinfo {author} {\bibfnamefont {J.-P.}\ \bibnamefont {Piquemal}}, \ and\ \bibinfo {author} {\bibfnamefont {P.}~\bibnamefont {Ren}},\ }\href {\doibase 10.1021/acs.jctc.9b00261} {\bibfield  {journal} {\bibinfo  {journal} {Journal of Chemical Theory and Computation}\ }\textbf {\bibinfo {volume} {15}},\ \bibinfo {pages} {4122} (\bibinfo {year} {2019})},\ \bibinfo {note} {pMID: 31136175},\ \Eprint {http://arxiv.org/abs/https://doi.org/10.1021/acs.jctc.9b00261} {https://doi.org/10.1021/acs.jctc.9b00261} \BibitemShut {NoStop}%
\bibitem [{\citenamefont {Liu}, \citenamefont {Piquemal},\ and\ \citenamefont {Ren}(2020)}]{AMOEBA+2}%
  \BibitemOpen
  \bibfield  {author} {\bibinfo {author} {\bibfnamefont {C.}~\bibnamefont {Liu}}, \bibinfo {author} {\bibfnamefont {J.-P.}\ \bibnamefont {Piquemal}}, \ and\ \bibinfo {author} {\bibfnamefont {P.}~\bibnamefont {Ren}},\ }\href {\doibase 10.1021/acs.jpclett.9b03489} {\bibfield  {journal} {\bibinfo  {journal} {The Journal of Physical Chemistry Letters}\ }\textbf {\bibinfo {volume} {11}},\ \bibinfo {pages} {419} (\bibinfo {year} {2020})},\ \bibinfo {note} {pMID: 31865706},\ \Eprint {http://arxiv.org/abs/https://doi.org/10.1021/acs.jpclett.9b03489} {https://doi.org/10.1021/acs.jpclett.9b03489} \BibitemShut {NoStop}%
\bibitem [{\citenamefont {Nawrocki}\ \emph {et~al.}(2022)\citenamefont {Nawrocki}, \citenamefont {Leontyev}, \citenamefont {Sakipov}, \citenamefont {Darkhovskiy}, \citenamefont {Kurnikov}, \citenamefont {Pereyaslavets}, \citenamefont {Kamath}, \citenamefont {Voronina}, \citenamefont {Butin}, \citenamefont {Illarionov} \emph {et~al.}}]{nawrocki2022protein}%
  \BibitemOpen
  \bibfield  {author} {\bibinfo {author} {\bibfnamefont {G.}~\bibnamefont {Nawrocki}}, \bibinfo {author} {\bibfnamefont {I.}~\bibnamefont {Leontyev}}, \bibinfo {author} {\bibfnamefont {S.}~\bibnamefont {Sakipov}}, \bibinfo {author} {\bibfnamefont {M.}~\bibnamefont {Darkhovskiy}}, \bibinfo {author} {\bibfnamefont {I.}~\bibnamefont {Kurnikov}}, \bibinfo {author} {\bibfnamefont {L.}~\bibnamefont {Pereyaslavets}}, \bibinfo {author} {\bibfnamefont {G.}~\bibnamefont {Kamath}}, \bibinfo {author} {\bibfnamefont {E.}~\bibnamefont {Voronina}}, \bibinfo {author} {\bibfnamefont {O.}~\bibnamefont {Butin}}, \bibinfo {author} {\bibfnamefont {A.}~\bibnamefont {Illarionov}},  \emph {et~al.},\ }\href@noop {} {\bibfield  {journal} {\bibinfo  {journal} {Journal of Chemical Theory and Computation}\ }\textbf {\bibinfo {volume} {18}},\ \bibinfo {pages} {7751} (\bibinfo {year} {2022})}\BibitemShut {NoStop}%
\bibitem [{\citenamefont {Melcr}\ and\ \citenamefont {Piquemal}(2019)}]{melcr2019accurate}%
  \BibitemOpen
  \bibfield  {author} {\bibinfo {author} {\bibfnamefont {J.}~\bibnamefont {Melcr}}\ and\ \bibinfo {author} {\bibfnamefont {J.-P.}\ \bibnamefont {Piquemal}},\ }\href@noop {} {\bibfield  {journal} {\bibinfo  {journal} {Frontiers in molecular biosciences}\ }\textbf {\bibinfo {volume} {6}},\ \bibinfo {pages} {143} (\bibinfo {year} {2019})}\BibitemShut {NoStop}%
\bibitem [{\citenamefont {Shi}\ \emph {et~al.}(2015)\citenamefont {Shi}, \citenamefont {Ren}, \citenamefont {Schnieders},\ and\ \citenamefont {Piquemal}}]{reviewcompchem}%
  \BibitemOpen
  \bibfield  {author} {\bibinfo {author} {\bibfnamefont {Y.}~\bibnamefont {Shi}}, \bibinfo {author} {\bibfnamefont {P.}~\bibnamefont {Ren}}, \bibinfo {author} {\bibfnamefont {M.}~\bibnamefont {Schnieders}}, \ and\ \bibinfo {author} {\bibfnamefont {J.-P.}\ \bibnamefont {Piquemal}},\ }\enquote {\bibinfo {title} {Polarizable force fields for biomolecular modeling},}\ in\ \href@noop {} {\emph {\bibinfo {booktitle} {Reviews in Computational Chemistry Volume 28}}}\ (\bibinfo  {publisher} {John Wiley and Sons, Ltd},\ \bibinfo {year} {2015})\ Chap.~\bibinfo {chapter} {2}, pp.\ \bibinfo {pages} {51--86}\BibitemShut {NoStop}%
\bibitem [{\citenamefont {Jing}\ \emph {et~al.}(2019)\citenamefont {Jing}, \citenamefont {Liu}, \citenamefont {Cheng}, \citenamefont {Qi}, \citenamefont {Walker}, \citenamefont {Piquemal},\ and\ \citenamefont {Ren}}]{annurev-biophys-070317-033349}%
  \BibitemOpen
  \bibfield  {author} {\bibinfo {author} {\bibfnamefont {Z.}~\bibnamefont {Jing}}, \bibinfo {author} {\bibfnamefont {C.}~\bibnamefont {Liu}}, \bibinfo {author} {\bibfnamefont {S.~Y.}\ \bibnamefont {Cheng}}, \bibinfo {author} {\bibfnamefont {R.}~\bibnamefont {Qi}}, \bibinfo {author} {\bibfnamefont {B.~D.}\ \bibnamefont {Walker}}, \bibinfo {author} {\bibfnamefont {J.-P.}\ \bibnamefont {Piquemal}}, \ and\ \bibinfo {author} {\bibfnamefont {P.}~\bibnamefont {Ren}},\ }\href@noop {} {\bibfield  {journal} {\bibinfo  {journal} {Ann. Rev. Biophys.}\ }\textbf {\bibinfo {volume} {48}},\ \bibinfo {pages} {371} (\bibinfo {year} {2019})}\BibitemShut {NoStop}%
\bibitem [{\citenamefont {Wang}\ \emph {et~al.}(2001)\citenamefont {Wang}, \citenamefont {Wang}, \citenamefont {Kollman},\ and\ \citenamefont {Case}}]{wang2001antechamber}%
  \BibitemOpen
  \bibfield  {author} {\bibinfo {author} {\bibfnamefont {J.}~\bibnamefont {Wang}}, \bibinfo {author} {\bibfnamefont {W.}~\bibnamefont {Wang}}, \bibinfo {author} {\bibfnamefont {P.~A.}\ \bibnamefont {Kollman}}, \ and\ \bibinfo {author} {\bibfnamefont {D.~A.}\ \bibnamefont {Case}},\ }\href@noop {} {\bibfield  {journal} {\bibinfo  {journal} {J. Am. Chem. Soc}\ }\textbf {\bibinfo {volume} {222}},\ \bibinfo {pages} {2001} (\bibinfo {year} {2001})}\BibitemShut {NoStop}%
\bibitem [{\citenamefont {Walker}\ \emph {et~al.}(2022)\citenamefont {Walker}, \citenamefont {Liu}, \citenamefont {Wait},\ and\ \citenamefont {Ren}}]{walker2022automation}%
  \BibitemOpen
  \bibfield  {author} {\bibinfo {author} {\bibfnamefont {B.}~\bibnamefont {Walker}}, \bibinfo {author} {\bibfnamefont {C.}~\bibnamefont {Liu}}, \bibinfo {author} {\bibfnamefont {E.}~\bibnamefont {Wait}}, \ and\ \bibinfo {author} {\bibfnamefont {P.}~\bibnamefont {Ren}},\ }\href@noop {} {\bibfield  {journal} {\bibinfo  {journal} {Journal of computational chemistry}\ }\textbf {\bibinfo {volume} {43}},\ \bibinfo {pages} {1530} (\bibinfo {year} {2022})}\BibitemShut {NoStop}%
\bibitem [{\citenamefont {Wang}\ \emph {et~al.}(2022)\citenamefont {Wang}, \citenamefont {Fass}, \citenamefont {Kaminow}, \citenamefont {Herr}, \citenamefont {Rufa}, \citenamefont {Zhang}, \citenamefont {Pulido}, \citenamefont {Henry}, \citenamefont {Macdonald}, \citenamefont {Takaba} \emph {et~al.}}]{wang2022end}%
  \BibitemOpen
  \bibfield  {author} {\bibinfo {author} {\bibfnamefont {Y.}~\bibnamefont {Wang}}, \bibinfo {author} {\bibfnamefont {J.}~\bibnamefont {Fass}}, \bibinfo {author} {\bibfnamefont {B.}~\bibnamefont {Kaminow}}, \bibinfo {author} {\bibfnamefont {J.~E.}\ \bibnamefont {Herr}}, \bibinfo {author} {\bibfnamefont {D.}~\bibnamefont {Rufa}}, \bibinfo {author} {\bibfnamefont {I.}~\bibnamefont {Zhang}}, \bibinfo {author} {\bibfnamefont {I.}~\bibnamefont {Pulido}}, \bibinfo {author} {\bibfnamefont {M.}~\bibnamefont {Henry}}, \bibinfo {author} {\bibfnamefont {H.~E.~B.}\ \bibnamefont {Macdonald}}, \bibinfo {author} {\bibfnamefont {K.}~\bibnamefont {Takaba}},  \emph {et~al.},\ }\href@noop {} {\bibfield  {journal} {\bibinfo  {journal} {Chemical Science}\ }\textbf {\bibinfo {volume} {13}},\ \bibinfo {pages} {12016} (\bibinfo {year} {2022})}\BibitemShut {NoStop}%
\bibitem [{\citenamefont {Chen}\ \emph {et~al.}(2023)\citenamefont {Chen}, \citenamefont {Inizan}, \citenamefont {Pl{\'e}}, \citenamefont {Lagardere}, \citenamefont {Piquemal},\ and\ \citenamefont {Maday}}]{chen2023advancing}%
  \BibitemOpen
  \bibfield  {author} {\bibinfo {author} {\bibfnamefont {G.}~\bibnamefont {Chen}}, \bibinfo {author} {\bibfnamefont {T.~J.}\ \bibnamefont {Inizan}}, \bibinfo {author} {\bibfnamefont {T.}~\bibnamefont {Pl{\'e}}}, \bibinfo {author} {\bibfnamefont {L.}~\bibnamefont {Lagardere}}, \bibinfo {author} {\bibfnamefont {J.-P.}\ \bibnamefont {Piquemal}}, \ and\ \bibinfo {author} {\bibfnamefont {Y.}~\bibnamefont {Maday}},\ }\href@noop {} {\  (\bibinfo {year} {2023})}\BibitemShut {NoStop}%
\bibitem [{\citenamefont {Thompson}\ \emph {et~al.}(2022)\citenamefont {Thompson}, \citenamefont {Aktulga}, \citenamefont {Berger}, \citenamefont {Bolintineanu}, \citenamefont {Brown}, \citenamefont {Crozier}, \citenamefont {In't~Veld}, \citenamefont {Kohlmeyer}, \citenamefont {Moore}, \citenamefont {Nguyen} \emph {et~al.}}]{thompson2022lammps}%
  \BibitemOpen
  \bibfield  {author} {\bibinfo {author} {\bibfnamefont {A.~P.}\ \bibnamefont {Thompson}}, \bibinfo {author} {\bibfnamefont {H.~M.}\ \bibnamefont {Aktulga}}, \bibinfo {author} {\bibfnamefont {R.}~\bibnamefont {Berger}}, \bibinfo {author} {\bibfnamefont {D.~S.}\ \bibnamefont {Bolintineanu}}, \bibinfo {author} {\bibfnamefont {W.~M.}\ \bibnamefont {Brown}}, \bibinfo {author} {\bibfnamefont {P.~S.}\ \bibnamefont {Crozier}}, \bibinfo {author} {\bibfnamefont {P.~J.}\ \bibnamefont {In't~Veld}}, \bibinfo {author} {\bibfnamefont {A.}~\bibnamefont {Kohlmeyer}}, \bibinfo {author} {\bibfnamefont {S.~G.}\ \bibnamefont {Moore}}, \bibinfo {author} {\bibfnamefont {T.~D.}\ \bibnamefont {Nguyen}},  \emph {et~al.},\ }\href@noop {} {\bibfield  {journal} {\bibinfo  {journal} {Computer Physics Communications}\ }\textbf {\bibinfo {volume} {271}},\ \bibinfo {pages} {108171} (\bibinfo {year} {2022})}\BibitemShut {NoStop}%
\bibitem [{\citenamefont {Case}\ \emph {et~al.}(2008)\citenamefont {Case}, \citenamefont {Darden}, \citenamefont {Cheatham}, \citenamefont {Simmerling}, \citenamefont {Wang}, \citenamefont {Duke}, \citenamefont {Luo}, \citenamefont {Crowley}, \citenamefont {Walker}, \citenamefont {Zhang} \emph {et~al.}}]{case2008amber}%
  \BibitemOpen
  \bibfield  {author} {\bibinfo {author} {\bibfnamefont {D.~A.}\ \bibnamefont {Case}}, \bibinfo {author} {\bibfnamefont {T.~A.}\ \bibnamefont {Darden}}, \bibinfo {author} {\bibfnamefont {T.~E.}\ \bibnamefont {Cheatham}}, \bibinfo {author} {\bibfnamefont {C.~L.}\ \bibnamefont {Simmerling}}, \bibinfo {author} {\bibfnamefont {J.}~\bibnamefont {Wang}}, \bibinfo {author} {\bibfnamefont {R.~E.}\ \bibnamefont {Duke}}, \bibinfo {author} {\bibfnamefont {R.}~\bibnamefont {Luo}}, \bibinfo {author} {\bibfnamefont {M.}~\bibnamefont {Crowley}}, \bibinfo {author} {\bibfnamefont {R.~C.}\ \bibnamefont {Walker}}, \bibinfo {author} {\bibfnamefont {W.}~\bibnamefont {Zhang}},  \emph {et~al.},\ }\href@noop {} {\  (\bibinfo {year} {2008})}\BibitemShut {NoStop}%
\bibitem [{\citenamefont {Van Der~Spoel}\ \emph {et~al.}(2005)\citenamefont {Van Der~Spoel}, \citenamefont {Lindahl}, \citenamefont {Hess}, \citenamefont {Groenhof}, \citenamefont {Mark},\ and\ \citenamefont {Berendsen}}]{van2005gromacs}%
  \BibitemOpen
  \bibfield  {author} {\bibinfo {author} {\bibfnamefont {D.}~\bibnamefont {Van Der~Spoel}}, \bibinfo {author} {\bibfnamefont {E.}~\bibnamefont {Lindahl}}, \bibinfo {author} {\bibfnamefont {B.}~\bibnamefont {Hess}}, \bibinfo {author} {\bibfnamefont {G.}~\bibnamefont {Groenhof}}, \bibinfo {author} {\bibfnamefont {A.~E.}\ \bibnamefont {Mark}}, \ and\ \bibinfo {author} {\bibfnamefont {H.~J.}\ \bibnamefont {Berendsen}},\ }\href@noop {} {\bibfield  {journal} {\bibinfo  {journal} {Journal of computational chemistry}\ }\textbf {\bibinfo {volume} {26}},\ \bibinfo {pages} {1701} (\bibinfo {year} {2005})}\BibitemShut {NoStop}%
\bibitem [{\citenamefont {Phillips}\ \emph {et~al.}(2005)\citenamefont {Phillips}, \citenamefont {Braun}, \citenamefont {Wang}, \citenamefont {Gumbart}, \citenamefont {Tajkhorshid}, \citenamefont {Villa}, \citenamefont {Chipot}, \citenamefont {Skeel}, \citenamefont {Kale},\ and\ \citenamefont {Schulten}}]{phillips2005scalable}%
  \BibitemOpen
  \bibfield  {author} {\bibinfo {author} {\bibfnamefont {J.~C.}\ \bibnamefont {Phillips}}, \bibinfo {author} {\bibfnamefont {R.}~\bibnamefont {Braun}}, \bibinfo {author} {\bibfnamefont {W.}~\bibnamefont {Wang}}, \bibinfo {author} {\bibfnamefont {J.}~\bibnamefont {Gumbart}}, \bibinfo {author} {\bibfnamefont {E.}~\bibnamefont {Tajkhorshid}}, \bibinfo {author} {\bibfnamefont {E.}~\bibnamefont {Villa}}, \bibinfo {author} {\bibfnamefont {C.}~\bibnamefont {Chipot}}, \bibinfo {author} {\bibfnamefont {R.~D.}\ \bibnamefont {Skeel}}, \bibinfo {author} {\bibfnamefont {L.}~\bibnamefont {Kale}}, \ and\ \bibinfo {author} {\bibfnamefont {K.}~\bibnamefont {Schulten}},\ }\href@noop {} {\bibfield  {journal} {\bibinfo  {journal} {Journal of computational chemistry}\ }\textbf {\bibinfo {volume} {26}},\ \bibinfo {pages} {1781} (\bibinfo {year} {2005})}\BibitemShut {NoStop}%
\bibitem [{\citenamefont {Lagardère}\ \emph {et~al.}(2018)\citenamefont {Lagardère}, \citenamefont {Jolly}, \citenamefont {Lipparini}, \citenamefont {Aviat}, \citenamefont {Stamm}, \citenamefont {Jing}, \citenamefont {Harger}, \citenamefont {Torabifard}, \citenamefont {Cisneros}, \citenamefont {Schnieders}, \citenamefont {Gresh}, \citenamefont {Maday}, \citenamefont {Ren}, \citenamefont {Ponder},\ and\ \citenamefont {Piquemal}}]{lagardere2018tinker}%
  \BibitemOpen
  \bibfield  {author} {\bibinfo {author} {\bibfnamefont {L.}~\bibnamefont {Lagardère}}, \bibinfo {author} {\bibfnamefont {L.-H.}\ \bibnamefont {Jolly}}, \bibinfo {author} {\bibfnamefont {F.}~\bibnamefont {Lipparini}}, \bibinfo {author} {\bibfnamefont {F.}~\bibnamefont {Aviat}}, \bibinfo {author} {\bibfnamefont {B.}~\bibnamefont {Stamm}}, \bibinfo {author} {\bibfnamefont {Z.~F.}\ \bibnamefont {Jing}}, \bibinfo {author} {\bibfnamefont {M.}~\bibnamefont {Harger}}, \bibinfo {author} {\bibfnamefont {H.}~\bibnamefont {Torabifard}}, \bibinfo {author} {\bibfnamefont {G.~A.}\ \bibnamefont {Cisneros}}, \bibinfo {author} {\bibfnamefont {M.~J.}\ \bibnamefont {Schnieders}}, \bibinfo {author} {\bibfnamefont {N.}~\bibnamefont {Gresh}}, \bibinfo {author} {\bibfnamefont {Y.}~\bibnamefont {Maday}}, \bibinfo {author} {\bibfnamefont {P.~Y.}\ \bibnamefont {Ren}}, \bibinfo {author} {\bibfnamefont {J.~W.}\ \bibnamefont {Ponder}}, \ and\ \bibinfo {author} {\bibfnamefont {J.-P.}\ \bibnamefont {Piquemal}},\ }\href {\doibase
  10.1039/C7SC04531J} {\bibfield  {journal} {\bibinfo  {journal} {Chemical Science}\ }\textbf {\bibinfo {volume} {9}},\ \bibinfo {pages} {956} (\bibinfo {year} {2018})}\BibitemShut {NoStop}%
\bibitem [{\citenamefont {Adjoua}\ \emph {et~al.}(2021)\citenamefont {Adjoua}, \citenamefont {Lagard{\`e}re}, \citenamefont {Jolly}, \citenamefont {Durocher}, \citenamefont {Very}, \citenamefont {Dupays}, \citenamefont {Wang}, \citenamefont {Jaffrellot~Inizan}, \citenamefont {C{\'e}lerse}, \citenamefont {Ren}, \citenamefont {Ponder},\ and\ \citenamefont {Piquemal}}]{adjoua2021tinker}%
  \BibitemOpen
  \bibfield  {author} {\bibinfo {author} {\bibfnamefont {O.}~\bibnamefont {Adjoua}}, \bibinfo {author} {\bibfnamefont {L.}~\bibnamefont {Lagard{\`e}re}}, \bibinfo {author} {\bibfnamefont {L.-H.}\ \bibnamefont {Jolly}}, \bibinfo {author} {\bibfnamefont {A.}~\bibnamefont {Durocher}}, \bibinfo {author} {\bibfnamefont {T.}~\bibnamefont {Very}}, \bibinfo {author} {\bibfnamefont {I.}~\bibnamefont {Dupays}}, \bibinfo {author} {\bibfnamefont {Z.}~\bibnamefont {Wang}}, \bibinfo {author} {\bibfnamefont {T.}~\bibnamefont {Jaffrellot~Inizan}}, \bibinfo {author} {\bibfnamefont {F.}~\bibnamefont {C{\'e}lerse}}, \bibinfo {author} {\bibfnamefont {P.}~\bibnamefont {Ren}}, \bibinfo {author} {\bibfnamefont {J.~W.}\ \bibnamefont {Ponder}}, \ and\ \bibinfo {author} {\bibfnamefont {J.-P.}\ \bibnamefont {Piquemal}},\ }\href@noop {} {\bibfield  {journal} {\bibinfo  {journal} {Journal of Chemical Theory and Computation}\ }\textbf {\bibinfo {volume} {17}},\ \bibinfo {pages} {2034} (\bibinfo {year} {2021})}\BibitemShut {NoStop}%
\bibitem [{\citenamefont {Chmiela}\ \emph {et~al.}(2018)\citenamefont {Chmiela}, \citenamefont {Sauceda}, \citenamefont {M{\"u}ller},\ and\ \citenamefont {Tkatchenko}}]{chmiela2018towards}%
  \BibitemOpen
  \bibfield  {author} {\bibinfo {author} {\bibfnamefont {S.}~\bibnamefont {Chmiela}}, \bibinfo {author} {\bibfnamefont {H.~E.}\ \bibnamefont {Sauceda}}, \bibinfo {author} {\bibfnamefont {K.-R.}\ \bibnamefont {M{\"u}ller}}, \ and\ \bibinfo {author} {\bibfnamefont {A.}~\bibnamefont {Tkatchenko}},\ }\href@noop {} {\bibfield  {journal} {\bibinfo  {journal} {Nature Communications}\ }\textbf {\bibinfo {volume} {9}},\ \bibinfo {pages} {1} (\bibinfo {year} {2018})}\BibitemShut {NoStop}%
\bibitem [{\citenamefont {Bigi}, \citenamefont {Pozdnyakov},\ and\ \citenamefont {Ceriotti}(2023)}]{bigi2023wigner}%
  \BibitemOpen
  \bibfield  {author} {\bibinfo {author} {\bibfnamefont {F.}~\bibnamefont {Bigi}}, \bibinfo {author} {\bibfnamefont {S.~N.}\ \bibnamefont {Pozdnyakov}}, \ and\ \bibinfo {author} {\bibfnamefont {M.}~\bibnamefont {Ceriotti}},\ }\href@noop {} {\bibfield  {journal} {\bibinfo  {journal} {arXiv preprint arXiv:2303.04124}\ } (\bibinfo {year} {2023})}\BibitemShut {NoStop}%
\bibitem [{\citenamefont {Shakouri}\ \emph {et~al.}(2017)\citenamefont {Shakouri}, \citenamefont {Behler}, \citenamefont {Meyer},\ and\ \citenamefont {Kroes}}]{shakouri2017accurate}%
  \BibitemOpen
  \bibfield  {author} {\bibinfo {author} {\bibfnamefont {K.}~\bibnamefont {Shakouri}}, \bibinfo {author} {\bibfnamefont {J.}~\bibnamefont {Behler}}, \bibinfo {author} {\bibfnamefont {J.}~\bibnamefont {Meyer}}, \ and\ \bibinfo {author} {\bibfnamefont {G.-J.}\ \bibnamefont {Kroes}},\ }\href@noop {} {\bibfield  {journal} {\bibinfo  {journal} {The journal of physical chemistry letters}\ }\textbf {\bibinfo {volume} {8}},\ \bibinfo {pages} {2131} (\bibinfo {year} {2017})}\BibitemShut {NoStop}%
\bibitem [{\citenamefont {Smith}, \citenamefont {Isayev},\ and\ \citenamefont {Roitberg}(2017)}]{smith2017ani}%
  \BibitemOpen
  \bibfield  {author} {\bibinfo {author} {\bibfnamefont {J.~S.}\ \bibnamefont {Smith}}, \bibinfo {author} {\bibfnamefont {O.}~\bibnamefont {Isayev}}, \ and\ \bibinfo {author} {\bibfnamefont {A.~E.}\ \bibnamefont {Roitberg}},\ }\href@noop {} {\bibfield  {journal} {\bibinfo  {journal} {Chemical Science}\ }\textbf {\bibinfo {volume} {8}},\ \bibinfo {pages} {3192} (\bibinfo {year} {2017})}\BibitemShut {NoStop}%
\bibitem [{\citenamefont {Sch{\"u}tt}\ \emph {et~al.}(2017)\citenamefont {Sch{\"u}tt}, \citenamefont {Kindermans}, \citenamefont {Sauceda~Felix}, \citenamefont {Chmiela}, \citenamefont {Tkatchenko},\ and\ \citenamefont {M{\"u}ller}}]{schutt2017schnet}%
  \BibitemOpen
  \bibfield  {author} {\bibinfo {author} {\bibfnamefont {K.}~\bibnamefont {Sch{\"u}tt}}, \bibinfo {author} {\bibfnamefont {P.-J.}\ \bibnamefont {Kindermans}}, \bibinfo {author} {\bibfnamefont {H.~E.}\ \bibnamefont {Sauceda~Felix}}, \bibinfo {author} {\bibfnamefont {S.}~\bibnamefont {Chmiela}}, \bibinfo {author} {\bibfnamefont {A.}~\bibnamefont {Tkatchenko}}, \ and\ \bibinfo {author} {\bibfnamefont {K.-R.}\ \bibnamefont {M{\"u}ller}},\ }\href@noop {} {\bibfield  {journal} {\bibinfo  {journal} {Advances in neural information processing systems}\ }\textbf {\bibinfo {volume} {30}} (\bibinfo {year} {2017})}\BibitemShut {NoStop}%
\bibitem [{\citenamefont {Lubbers}, \citenamefont {Smith},\ and\ \citenamefont {Barros}(2018)}]{lubbers2018hierarchical}%
  \BibitemOpen
  \bibfield  {author} {\bibinfo {author} {\bibfnamefont {N.}~\bibnamefont {Lubbers}}, \bibinfo {author} {\bibfnamefont {J.~S.}\ \bibnamefont {Smith}}, \ and\ \bibinfo {author} {\bibfnamefont {K.}~\bibnamefont {Barros}},\ }\href@noop {} {\bibfield  {journal} {\bibinfo  {journal} {The Journal of Chemical Physics}\ }\textbf {\bibinfo {volume} {148}} (\bibinfo {year} {2018})}\BibitemShut {NoStop}%
\bibitem [{\citenamefont {Zubatyuk}\ \emph {et~al.}(2019)\citenamefont {Zubatyuk}, \citenamefont {Smith}, \citenamefont {Leszczynski},\ and\ \citenamefont {Isayev}}]{zubatyuk2019accurate}%
  \BibitemOpen
  \bibfield  {author} {\bibinfo {author} {\bibfnamefont {R.}~\bibnamefont {Zubatyuk}}, \bibinfo {author} {\bibfnamefont {J.~S.}\ \bibnamefont {Smith}}, \bibinfo {author} {\bibfnamefont {J.}~\bibnamefont {Leszczynski}}, \ and\ \bibinfo {author} {\bibfnamefont {O.}~\bibnamefont {Isayev}},\ }\href@noop {} {\bibfield  {journal} {\bibinfo  {journal} {Science advances}\ }\textbf {\bibinfo {volume} {5}},\ \bibinfo {pages} {eaav6490} (\bibinfo {year} {2019})}\BibitemShut {NoStop}%
\bibitem [{\citenamefont {Qiao}\ \emph {et~al.}(2021)\citenamefont {Qiao}, \citenamefont {Christensen}, \citenamefont {Welborn}, \citenamefont {Manby}, \citenamefont {Anandkumar},\ and\ \citenamefont {Miller~III}}]{qiao2021unite}%
  \BibitemOpen
  \bibfield  {author} {\bibinfo {author} {\bibfnamefont {Z.}~\bibnamefont {Qiao}}, \bibinfo {author} {\bibfnamefont {A.~S.}\ \bibnamefont {Christensen}}, \bibinfo {author} {\bibfnamefont {M.}~\bibnamefont {Welborn}}, \bibinfo {author} {\bibfnamefont {F.~R.}\ \bibnamefont {Manby}}, \bibinfo {author} {\bibfnamefont {A.}~\bibnamefont {Anandkumar}}, \ and\ \bibinfo {author} {\bibfnamefont {T.~F.}\ \bibnamefont {Miller~III}},\ }\href@noop {} {\bibfield  {journal} {\bibinfo  {journal} {arXiv preprint arXiv:2105.14655}\ } (\bibinfo {year} {2021})}\BibitemShut {NoStop}%
\bibitem [{\citenamefont {Unke}\ \emph {et~al.}(2021)\citenamefont {Unke}, \citenamefont {Chmiela}, \citenamefont {Gastegger}, \citenamefont {Sch{\"u}tt}, \citenamefont {Sauceda},\ and\ \citenamefont {M{\"u}ller}}]{unke2021spookynet}%
  \BibitemOpen
  \bibfield  {author} {\bibinfo {author} {\bibfnamefont {O.~T.}\ \bibnamefont {Unke}}, \bibinfo {author} {\bibfnamefont {S.}~\bibnamefont {Chmiela}}, \bibinfo {author} {\bibfnamefont {M.}~\bibnamefont {Gastegger}}, \bibinfo {author} {\bibfnamefont {K.~T.}\ \bibnamefont {Sch{\"u}tt}}, \bibinfo {author} {\bibfnamefont {H.~E.}\ \bibnamefont {Sauceda}}, \ and\ \bibinfo {author} {\bibfnamefont {K.-R.}\ \bibnamefont {M{\"u}ller}},\ }\href@noop {} {\bibfield  {journal} {\bibinfo  {journal} {Nature Communications}\ }\textbf {\bibinfo {volume} {12}},\ \bibinfo {pages} {1} (\bibinfo {year} {2021})}\BibitemShut {NoStop}%
\bibitem [{\citenamefont {Gasteiger}, \citenamefont {Becker},\ and\ \citenamefont {G{\"u}nnemann}(2021)}]{gasteiger2021gemnet}%
  \BibitemOpen
  \bibfield  {author} {\bibinfo {author} {\bibfnamefont {J.}~\bibnamefont {Gasteiger}}, \bibinfo {author} {\bibfnamefont {F.}~\bibnamefont {Becker}}, \ and\ \bibinfo {author} {\bibfnamefont {S.}~\bibnamefont {G{\"u}nnemann}},\ }\href@noop {} {\bibfield  {journal} {\bibinfo  {journal} {Advances in Neural Information Processing Systems}\ }\textbf {\bibinfo {volume} {34}},\ \bibinfo {pages} {6790} (\bibinfo {year} {2021})}\BibitemShut {NoStop}%
\bibitem [{\citenamefont {Batzner}\ \emph {et~al.}(2022)\citenamefont {Batzner}, \citenamefont {Musaelian}, \citenamefont {Sun}, \citenamefont {Geiger}, \citenamefont {Mailoa}, \citenamefont {Kornbluth}, \citenamefont {Molinari}, \citenamefont {Smidt},\ and\ \citenamefont {Kozinsky}}]{batzner20223}%
  \BibitemOpen
  \bibfield  {author} {\bibinfo {author} {\bibfnamefont {S.}~\bibnamefont {Batzner}}, \bibinfo {author} {\bibfnamefont {A.}~\bibnamefont {Musaelian}}, \bibinfo {author} {\bibfnamefont {L.}~\bibnamefont {Sun}}, \bibinfo {author} {\bibfnamefont {M.}~\bibnamefont {Geiger}}, \bibinfo {author} {\bibfnamefont {J.~P.}\ \bibnamefont {Mailoa}}, \bibinfo {author} {\bibfnamefont {M.}~\bibnamefont {Kornbluth}}, \bibinfo {author} {\bibfnamefont {N.}~\bibnamefont {Molinari}}, \bibinfo {author} {\bibfnamefont {T.~E.}\ \bibnamefont {Smidt}}, \ and\ \bibinfo {author} {\bibfnamefont {B.}~\bibnamefont {Kozinsky}},\ }\href@noop {} {\bibfield  {journal} {\bibinfo  {journal} {Nature Communications}\ }\textbf {\bibinfo {volume} {13}},\ \bibinfo {pages} {1} (\bibinfo {year} {2022})}\BibitemShut {NoStop}%
\bibitem [{\citenamefont {Musaelian}\ \emph {et~al.}(2023)\citenamefont {Musaelian}, \citenamefont {Batzner}, \citenamefont {Johansson}, \citenamefont {Sun}, \citenamefont {Owen}, \citenamefont {Kornbluth},\ and\ \citenamefont {Kozinsky}}]{musaelian2022learning}%
  \BibitemOpen
  \bibfield  {author} {\bibinfo {author} {\bibfnamefont {A.}~\bibnamefont {Musaelian}}, \bibinfo {author} {\bibfnamefont {S.}~\bibnamefont {Batzner}}, \bibinfo {author} {\bibfnamefont {A.}~\bibnamefont {Johansson}}, \bibinfo {author} {\bibfnamefont {L.}~\bibnamefont {Sun}}, \bibinfo {author} {\bibfnamefont {C.~J.}\ \bibnamefont {Owen}}, \bibinfo {author} {\bibfnamefont {M.}~\bibnamefont {Kornbluth}}, \ and\ \bibinfo {author} {\bibfnamefont {B.}~\bibnamefont {Kozinsky}},\ }\href@noop {} {\bibfield  {journal} {\bibinfo  {journal} {Nature Communications}\ }\textbf {\bibinfo {volume} {14}},\ \bibinfo {pages} {579} (\bibinfo {year} {2023})}\BibitemShut {NoStop}%
\bibitem [{\citenamefont {Grisafi}\ and\ \citenamefont {Ceriotti}(2019)}]{grisafi2019incorporating}%
  \BibitemOpen
  \bibfield  {author} {\bibinfo {author} {\bibfnamefont {A.}~\bibnamefont {Grisafi}}\ and\ \bibinfo {author} {\bibfnamefont {M.}~\bibnamefont {Ceriotti}},\ }\href@noop {} {\bibfield  {journal} {\bibinfo  {journal} {The Journal of Chemical Physics}\ }\textbf {\bibinfo {volume} {151}},\ \bibinfo {pages} {204105} (\bibinfo {year} {2019})}\BibitemShut {NoStop}%
\bibitem [{\citenamefont {Grisafi}, \citenamefont {Nigam},\ and\ \citenamefont {Ceriotti}(2021)}]{grisafi2021multi}%
  \BibitemOpen
  \bibfield  {author} {\bibinfo {author} {\bibfnamefont {A.}~\bibnamefont {Grisafi}}, \bibinfo {author} {\bibfnamefont {J.}~\bibnamefont {Nigam}}, \ and\ \bibinfo {author} {\bibfnamefont {M.}~\bibnamefont {Ceriotti}},\ }\href@noop {} {\bibfield  {journal} {\bibinfo  {journal} {Chemical Science}\ }\textbf {\bibinfo {volume} {12}},\ \bibinfo {pages} {2078} (\bibinfo {year} {2021})}\BibitemShut {NoStop}%
\bibitem [{\citenamefont {Ko}\ \emph {et~al.}(2021)\citenamefont {Ko}, \citenamefont {Finkler}, \citenamefont {Goedecker},\ and\ \citenamefont {Behler}}]{ko2021fourth}%
  \BibitemOpen
  \bibfield  {author} {\bibinfo {author} {\bibfnamefont {T.~W.}\ \bibnamefont {Ko}}, \bibinfo {author} {\bibfnamefont {J.~A.}\ \bibnamefont {Finkler}}, \bibinfo {author} {\bibfnamefont {S.}~\bibnamefont {Goedecker}}, \ and\ \bibinfo {author} {\bibfnamefont {J.}~\bibnamefont {Behler}},\ }\href@noop {} {\bibfield  {journal} {\bibinfo  {journal} {Nature Communications}\ }\textbf {\bibinfo {volume} {12}},\ \bibinfo {pages} {1} (\bibinfo {year} {2021})}\BibitemShut {NoStop}%
\bibitem [{\citenamefont {Unke}\ and\ \citenamefont {Meuwly}(2019)}]{unke2019physnet}%
  \BibitemOpen
  \bibfield  {author} {\bibinfo {author} {\bibfnamefont {O.~T.}\ \bibnamefont {Unke}}\ and\ \bibinfo {author} {\bibfnamefont {M.}~\bibnamefont {Meuwly}},\ }\href@noop {} {\bibfield  {journal} {\bibinfo  {journal} {Journal of Chemical Theory and Computation}\ }\textbf {\bibinfo {volume} {15}},\ \bibinfo {pages} {3678} (\bibinfo {year} {2019})}\BibitemShut {NoStop}%
\bibitem [{\citenamefont {Qu}\ \emph {et~al.}(2023)\citenamefont {Qu}, \citenamefont {Yu}, \citenamefont {Houston}, \citenamefont {Conte}, \citenamefont {Nandi},\ and\ \citenamefont {Bowman}}]{qu2023interfacing}%
  \BibitemOpen
  \bibfield  {author} {\bibinfo {author} {\bibfnamefont {C.}~\bibnamefont {Qu}}, \bibinfo {author} {\bibfnamefont {Q.}~\bibnamefont {Yu}}, \bibinfo {author} {\bibfnamefont {P.~L.}\ \bibnamefont {Houston}}, \bibinfo {author} {\bibfnamefont {R.}~\bibnamefont {Conte}}, \bibinfo {author} {\bibfnamefont {A.}~\bibnamefont {Nandi}}, \ and\ \bibinfo {author} {\bibfnamefont {J.~M.}\ \bibnamefont {Bowman}},\ }\href@noop {} {\bibfield  {journal} {\bibinfo  {journal} {Journal of Chemical Theory and Computation}\ } (\bibinfo {year} {2023})}\BibitemShut {NoStop}%
\bibitem [{\citenamefont {Yang}\ \emph {et~al.}(2022)\citenamefont {Yang}, \citenamefont {Li}, \citenamefont {Chen},\ and\ \citenamefont {Yu}}]{yang2022transferrable}%
  \BibitemOpen
  \bibfield  {author} {\bibinfo {author} {\bibfnamefont {L.}~\bibnamefont {Yang}}, \bibinfo {author} {\bibfnamefont {J.}~\bibnamefont {Li}}, \bibinfo {author} {\bibfnamefont {F.}~\bibnamefont {Chen}}, \ and\ \bibinfo {author} {\bibfnamefont {K.}~\bibnamefont {Yu}},\ }\href@noop {} {\bibfield  {journal} {\bibinfo  {journal} {The Journal of Chemical Physics}\ }\textbf {\bibinfo {volume} {157}},\ \bibinfo {pages} {214108} (\bibinfo {year} {2022})}\BibitemShut {NoStop}%
\bibitem [{\citenamefont {Zhang}\ \emph {et~al.}(2022)\citenamefont {Zhang}, \citenamefont {Wang}, \citenamefont {Muniz}, \citenamefont {Panagiotopoulos}, \citenamefont {Car},\ and\ \citenamefont {E}}]{zhang2022deep}%
  \BibitemOpen
  \bibfield  {author} {\bibinfo {author} {\bibfnamefont {L.}~\bibnamefont {Zhang}}, \bibinfo {author} {\bibfnamefont {H.}~\bibnamefont {Wang}}, \bibinfo {author} {\bibfnamefont {M.~C.}\ \bibnamefont {Muniz}}, \bibinfo {author} {\bibfnamefont {A.~Z.}\ \bibnamefont {Panagiotopoulos}}, \bibinfo {author} {\bibfnamefont {R.}~\bibnamefont {Car}}, \ and\ \bibinfo {author} {\bibfnamefont {W.}~\bibnamefont {E}},\ }\href@noop {} {\bibfield  {journal} {\bibinfo  {journal} {The Journal of Chemical Physics}\ }\textbf {\bibinfo {volume} {156}},\ \bibinfo {pages} {124107} (\bibinfo {year} {2022})}\BibitemShut {NoStop}%
\bibitem [{\citenamefont {Bowman}\ \emph {et~al.}(2022)\citenamefont {Bowman}, \citenamefont {Qu}, \citenamefont {Conte}, \citenamefont {Nandi}, \citenamefont {Houston},\ and\ \citenamefont {Yu}}]{bowman2022delta}%
  \BibitemOpen
  \bibfield  {author} {\bibinfo {author} {\bibfnamefont {J.~M.}\ \bibnamefont {Bowman}}, \bibinfo {author} {\bibfnamefont {C.}~\bibnamefont {Qu}}, \bibinfo {author} {\bibfnamefont {R.}~\bibnamefont {Conte}}, \bibinfo {author} {\bibfnamefont {A.}~\bibnamefont {Nandi}}, \bibinfo {author} {\bibfnamefont {P.~L.}\ \bibnamefont {Houston}}, \ and\ \bibinfo {author} {\bibfnamefont {Q.}~\bibnamefont {Yu}},\ }\href@noop {} {\bibfield  {journal} {\bibinfo  {journal} {Journal of Chemical Theory and Computation}\ }\textbf {\bibinfo {volume} {19}},\ \bibinfo {pages} {1} (\bibinfo {year} {2022})}\BibitemShut {NoStop}%
\bibitem [{\citenamefont {Chen}\ and\ \citenamefont {Yu}(2023)}]{chen2023phyneo}%
  \BibitemOpen
  \bibfield  {author} {\bibinfo {author} {\bibfnamefont {J.}~\bibnamefont {Chen}}\ and\ \bibinfo {author} {\bibfnamefont {K.}~\bibnamefont {Yu}},\ }\href@noop {} {\bibfield  {journal} {\bibinfo  {journal} {Journal of Chemical Theory and Computation}\ }\textbf {\bibinfo {volume} {20}},\ \bibinfo {pages} {253} (\bibinfo {year} {2023})}\BibitemShut {NoStop}%
\bibitem [{\citenamefont {Wang}\ \emph {et~al.}(2024{\natexlab{a}})\citenamefont {Wang}, \citenamefont {Inizan}, \citenamefont {Liu}, \citenamefont {Piquemal},\ and\ \citenamefont {Ren}}]{AMOEBA-NN}%
  \BibitemOpen
  \bibfield  {author} {\bibinfo {author} {\bibfnamefont {Y.}~\bibnamefont {Wang}}, \bibinfo {author} {\bibfnamefont {T.~J.}\ \bibnamefont {Inizan}}, \bibinfo {author} {\bibfnamefont {C.}~\bibnamefont {Liu}}, \bibinfo {author} {\bibfnamefont {J.-P.}\ \bibnamefont {Piquemal}}, \ and\ \bibinfo {author} {\bibfnamefont {P.}~\bibnamefont {Ren}},\ }\href {\doibase 10.1021/acs.jpcb.3c08166} {\bibfield  {journal} {\bibinfo  {journal} {The Journal of Physical Chemistry B}\ }\textbf {\bibinfo {volume} {128}},\ \bibinfo {pages} {2381} (\bibinfo {year} {2024}{\natexlab{a}})},\ \bibinfo {note} {pMID: 38445577},\ \Eprint {http://arxiv.org/abs/https://doi.org/10.1021/acs.jpcb.3c08166} {https://doi.org/10.1021/acs.jpcb.3c08166} \BibitemShut {NoStop}%
\bibitem [{\citenamefont {Plé}, \citenamefont {Lagardère},\ and\ \citenamefont {Piquemal}(2023)}]{ple2023fennix}%
  \BibitemOpen
  \bibfield  {author} {\bibinfo {author} {\bibfnamefont {T.}~\bibnamefont {Plé}}, \bibinfo {author} {\bibfnamefont {L.}~\bibnamefont {Lagardère}}, \ and\ \bibinfo {author} {\bibfnamefont {J.-P.}\ \bibnamefont {Piquemal}},\ }\href {\doibase 10.1039/D3SC02581K} {\bibfield  {journal} {\bibinfo  {journal} {Chemical Science}\ }\textbf {\bibinfo {volume} {14}},\ \bibinfo {pages} {12554} (\bibinfo {year} {2023})}\BibitemShut {NoStop}%
\bibitem [{\citenamefont {Sch{\"u}tt}\ \emph {et~al.}(2023)\citenamefont {Sch{\"u}tt}, \citenamefont {Hessmann}, \citenamefont {Gebauer}, \citenamefont {Lederer},\ and\ \citenamefont {Gastegger}}]{schutt2023schnetpack}%
  \BibitemOpen
  \bibfield  {author} {\bibinfo {author} {\bibfnamefont {K.~T.}\ \bibnamefont {Sch{\"u}tt}}, \bibinfo {author} {\bibfnamefont {S.~S.}\ \bibnamefont {Hessmann}}, \bibinfo {author} {\bibfnamefont {N.~W.}\ \bibnamefont {Gebauer}}, \bibinfo {author} {\bibfnamefont {J.}~\bibnamefont {Lederer}}, \ and\ \bibinfo {author} {\bibfnamefont {M.}~\bibnamefont {Gastegger}},\ }\href@noop {} {\bibfield  {journal} {\bibinfo  {journal} {The Journal of Chemical Physics}\ }\textbf {\bibinfo {volume} {158}} (\bibinfo {year} {2023})}\BibitemShut {NoStop}%
\bibitem [{\citenamefont {Dral}\ \emph {et~al.}(2024)\citenamefont {Dral}, \citenamefont {Ge}, \citenamefont {Hou}, \citenamefont {Zheng}, \citenamefont {Chen}, \citenamefont {Barbatti}, \citenamefont {Isayev}, \citenamefont {Wang}, \citenamefont {Xue}, \citenamefont {Pinheiro~Jr} \emph {et~al.}}]{dral2024mlatom}%
  \BibitemOpen
  \bibfield  {author} {\bibinfo {author} {\bibfnamefont {P.~O.}\ \bibnamefont {Dral}}, \bibinfo {author} {\bibfnamefont {F.}~\bibnamefont {Ge}}, \bibinfo {author} {\bibfnamefont {Y.-F.}\ \bibnamefont {Hou}}, \bibinfo {author} {\bibfnamefont {P.}~\bibnamefont {Zheng}}, \bibinfo {author} {\bibfnamefont {Y.}~\bibnamefont {Chen}}, \bibinfo {author} {\bibfnamefont {M.}~\bibnamefont {Barbatti}}, \bibinfo {author} {\bibfnamefont {O.}~\bibnamefont {Isayev}}, \bibinfo {author} {\bibfnamefont {C.}~\bibnamefont {Wang}}, \bibinfo {author} {\bibfnamefont {B.-X.}\ \bibnamefont {Xue}}, \bibinfo {author} {\bibfnamefont {M.}~\bibnamefont {Pinheiro~Jr}},  \emph {et~al.},\ }\href@noop {} {\bibfield  {journal} {\bibinfo  {journal} {Journal of Chemical Theory and Computation}\ } (\bibinfo {year} {2024})}\BibitemShut {NoStop}%
\bibitem [{\citenamefont {Zeng}\ \emph {et~al.}(2023)\citenamefont {Zeng}, \citenamefont {Zhang}, \citenamefont {Lu}, \citenamefont {Mo}, \citenamefont {Li}, \citenamefont {Chen}, \citenamefont {Rynik}, \citenamefont {Huang}, \citenamefont {Li}, \citenamefont {Shi} \emph {et~al.}}]{zeng2023deepmd}%
  \BibitemOpen
  \bibfield  {author} {\bibinfo {author} {\bibfnamefont {J.}~\bibnamefont {Zeng}}, \bibinfo {author} {\bibfnamefont {D.}~\bibnamefont {Zhang}}, \bibinfo {author} {\bibfnamefont {D.}~\bibnamefont {Lu}}, \bibinfo {author} {\bibfnamefont {P.}~\bibnamefont {Mo}}, \bibinfo {author} {\bibfnamefont {Z.}~\bibnamefont {Li}}, \bibinfo {author} {\bibfnamefont {Y.}~\bibnamefont {Chen}}, \bibinfo {author} {\bibfnamefont {M.}~\bibnamefont {Rynik}}, \bibinfo {author} {\bibfnamefont {L.}~\bibnamefont {Huang}}, \bibinfo {author} {\bibfnamefont {Z.}~\bibnamefont {Li}}, \bibinfo {author} {\bibfnamefont {S.}~\bibnamefont {Shi}},  \emph {et~al.},\ }\href@noop {} {\bibfield  {journal} {\bibinfo  {journal} {The Journal of Chemical Physics}\ }\textbf {\bibinfo {volume} {159}} (\bibinfo {year} {2023})}\BibitemShut {NoStop}%
\bibitem [{\citenamefont {Bereau}, \citenamefont {Andrienko},\ and\ \citenamefont {Von~Lilienfeld}(2015)}]{bereau2015transferable}%
  \BibitemOpen
  \bibfield  {author} {\bibinfo {author} {\bibfnamefont {T.}~\bibnamefont {Bereau}}, \bibinfo {author} {\bibfnamefont {D.}~\bibnamefont {Andrienko}}, \ and\ \bibinfo {author} {\bibfnamefont {O.~A.}\ \bibnamefont {Von~Lilienfeld}},\ }\href@noop {} {\bibfield  {journal} {\bibinfo  {journal} {Journal of chemical theory and computation}\ }\textbf {\bibinfo {volume} {11}},\ \bibinfo {pages} {3225} (\bibinfo {year} {2015})}\BibitemShut {NoStop}%
\bibitem [{\citenamefont {Glick}\ \emph {et~al.}(2021)\citenamefont {Glick}, \citenamefont {Koutsoukas}, \citenamefont {Cheney},\ and\ \citenamefont {Sherrill}}]{glick2021cartesian}%
  \BibitemOpen
  \bibfield  {author} {\bibinfo {author} {\bibfnamefont {Z.~L.}\ \bibnamefont {Glick}}, \bibinfo {author} {\bibfnamefont {A.}~\bibnamefont {Koutsoukas}}, \bibinfo {author} {\bibfnamefont {D.~L.}\ \bibnamefont {Cheney}}, \ and\ \bibinfo {author} {\bibfnamefont {C.~D.}\ \bibnamefont {Sherrill}},\ }\href@noop {} {\bibfield  {journal} {\bibinfo  {journal} {The Journal of Chemical Physics}\ }\textbf {\bibinfo {volume} {154}} (\bibinfo {year} {2021})}\BibitemShut {NoStop}%
\bibitem [{\citenamefont {Thurlemann}, \citenamefont {Boselt},\ and\ \citenamefont {Riniker}(2022)}]{thurlemann2022learning}%
  \BibitemOpen
  \bibfield  {author} {\bibinfo {author} {\bibfnamefont {M.}~\bibnamefont {Thurlemann}}, \bibinfo {author} {\bibfnamefont {L.}~\bibnamefont {Boselt}}, \ and\ \bibinfo {author} {\bibfnamefont {S.}~\bibnamefont {Riniker}},\ }\href@noop {} {\bibfield  {journal} {\bibinfo  {journal} {Journal of Chemical Theory and Computation}\ }\textbf {\bibinfo {volume} {18}},\ \bibinfo {pages} {1701} (\bibinfo {year} {2022})}\BibitemShut {NoStop}%
\bibitem [{\citenamefont {Bradbury}\ \emph {et~al.}(2018)\citenamefont {Bradbury}, \citenamefont {Frostig}, \citenamefont {Hawkins}, \citenamefont {Johnson}, \citenamefont {Leary}, \citenamefont {Maclaurin}, \citenamefont {Necula}, \citenamefont {Paszke}, \citenamefont {Vander{P}las}, \citenamefont {Wanderman-{M}ilne},\ and\ \citenamefont {Zhang}}]{jax2018github}%
  \BibitemOpen
  \bibfield  {author} {\bibinfo {author} {\bibfnamefont {J.}~\bibnamefont {Bradbury}}, \bibinfo {author} {\bibfnamefont {R.}~\bibnamefont {Frostig}}, \bibinfo {author} {\bibfnamefont {P.}~\bibnamefont {Hawkins}}, \bibinfo {author} {\bibfnamefont {M.~J.}\ \bibnamefont {Johnson}}, \bibinfo {author} {\bibfnamefont {C.}~\bibnamefont {Leary}}, \bibinfo {author} {\bibfnamefont {D.}~\bibnamefont {Maclaurin}}, \bibinfo {author} {\bibfnamefont {G.}~\bibnamefont {Necula}}, \bibinfo {author} {\bibfnamefont {A.}~\bibnamefont {Paszke}}, \bibinfo {author} {\bibfnamefont {J.}~\bibnamefont {Vander{P}las}}, \bibinfo {author} {\bibfnamefont {S.}~\bibnamefont {Wanderman-{M}ilne}}, \ and\ \bibinfo {author} {\bibfnamefont {Q.}~\bibnamefont {Zhang}},\ }\href {http://github.com/google/jax} {\enquote {\bibinfo {title} {{JAX}: composable transformations of {P}ython+{N}um{P}y programs},}\ } (\bibinfo {year} {2018})\BibitemShut {NoStop}%
\bibitem [{\citenamefont {DeepMind}\ \emph {et~al.}(2020)\citenamefont {DeepMind}, \citenamefont {Babuschkin}, \citenamefont {Baumli}, \citenamefont {Bell}, \citenamefont {Bhupatiraju}, \citenamefont {Bruce}, \citenamefont {Buchlovsky}, \citenamefont {Budden}, \citenamefont {Cai}, \citenamefont {Clark}, \citenamefont {Danihelka}, \citenamefont {Dedieu}, \citenamefont {Fantacci}, \citenamefont {Godwin}, \citenamefont {Jones}, \citenamefont {Hemsley}, \citenamefont {Hennigan}, \citenamefont {Hessel}, \citenamefont {Hou}, \citenamefont {Kapturowski}, \citenamefont {Keck}, \citenamefont {Kemaev}, \citenamefont {King}, \citenamefont {Kunesch}, \citenamefont {Martens}, \citenamefont {Merzic}, \citenamefont {Mikulik}, \citenamefont {Norman}, \citenamefont {Papamakarios}, \citenamefont {Quan}, \citenamefont {Ring}, \citenamefont {Ruiz}, \citenamefont {Sanchez}, \citenamefont {Sartran}, \citenamefont {Schneider}, \citenamefont {Sezener}, \citenamefont {Spencer}, \citenamefont {Srinivasan}, \citenamefont
  {Stanojevi\'{c}}, \citenamefont {Stokowiec}, \citenamefont {Wang}, \citenamefont {Zhou},\ and\ \citenamefont {Viola}}]{deepmind2020jax}%
  \BibitemOpen
  \bibfield  {author} {\bibinfo {author} {\bibnamefont {DeepMind}}, \bibinfo {author} {\bibfnamefont {I.}~\bibnamefont {Babuschkin}}, \bibinfo {author} {\bibfnamefont {K.}~\bibnamefont {Baumli}}, \bibinfo {author} {\bibfnamefont {A.}~\bibnamefont {Bell}}, \bibinfo {author} {\bibfnamefont {S.}~\bibnamefont {Bhupatiraju}}, \bibinfo {author} {\bibfnamefont {J.}~\bibnamefont {Bruce}}, \bibinfo {author} {\bibfnamefont {P.}~\bibnamefont {Buchlovsky}}, \bibinfo {author} {\bibfnamefont {D.}~\bibnamefont {Budden}}, \bibinfo {author} {\bibfnamefont {T.}~\bibnamefont {Cai}}, \bibinfo {author} {\bibfnamefont {A.}~\bibnamefont {Clark}}, \bibinfo {author} {\bibfnamefont {I.}~\bibnamefont {Danihelka}}, \bibinfo {author} {\bibfnamefont {A.}~\bibnamefont {Dedieu}}, \bibinfo {author} {\bibfnamefont {C.}~\bibnamefont {Fantacci}}, \bibinfo {author} {\bibfnamefont {J.}~\bibnamefont {Godwin}}, \bibinfo {author} {\bibfnamefont {C.}~\bibnamefont {Jones}}, \bibinfo {author} {\bibfnamefont {R.}~\bibnamefont {Hemsley}}, \bibinfo
  {author} {\bibfnamefont {T.}~\bibnamefont {Hennigan}}, \bibinfo {author} {\bibfnamefont {M.}~\bibnamefont {Hessel}}, \bibinfo {author} {\bibfnamefont {S.}~\bibnamefont {Hou}}, \bibinfo {author} {\bibfnamefont {S.}~\bibnamefont {Kapturowski}}, \bibinfo {author} {\bibfnamefont {T.}~\bibnamefont {Keck}}, \bibinfo {author} {\bibfnamefont {I.}~\bibnamefont {Kemaev}}, \bibinfo {author} {\bibfnamefont {M.}~\bibnamefont {King}}, \bibinfo {author} {\bibfnamefont {M.}~\bibnamefont {Kunesch}}, \bibinfo {author} {\bibfnamefont {L.}~\bibnamefont {Martens}}, \bibinfo {author} {\bibfnamefont {H.}~\bibnamefont {Merzic}}, \bibinfo {author} {\bibfnamefont {V.}~\bibnamefont {Mikulik}}, \bibinfo {author} {\bibfnamefont {T.}~\bibnamefont {Norman}}, \bibinfo {author} {\bibfnamefont {G.}~\bibnamefont {Papamakarios}}, \bibinfo {author} {\bibfnamefont {J.}~\bibnamefont {Quan}}, \bibinfo {author} {\bibfnamefont {R.}~\bibnamefont {Ring}}, \bibinfo {author} {\bibfnamefont {F.}~\bibnamefont {Ruiz}}, \bibinfo {author} {\bibfnamefont
  {A.}~\bibnamefont {Sanchez}}, \bibinfo {author} {\bibfnamefont {L.}~\bibnamefont {Sartran}}, \bibinfo {author} {\bibfnamefont {R.}~\bibnamefont {Schneider}}, \bibinfo {author} {\bibfnamefont {E.}~\bibnamefont {Sezener}}, \bibinfo {author} {\bibfnamefont {S.}~\bibnamefont {Spencer}}, \bibinfo {author} {\bibfnamefont {S.}~\bibnamefont {Srinivasan}}, \bibinfo {author} {\bibfnamefont {M.}~\bibnamefont {Stanojevi\'{c}}}, \bibinfo {author} {\bibfnamefont {W.}~\bibnamefont {Stokowiec}}, \bibinfo {author} {\bibfnamefont {L.}~\bibnamefont {Wang}}, \bibinfo {author} {\bibfnamefont {G.}~\bibnamefont {Zhou}}, \ and\ \bibinfo {author} {\bibfnamefont {F.}~\bibnamefont {Viola}},\ }\href {http://github.com/google-deepmind} {\enquote {\bibinfo {title} {The {D}eep{M}ind {JAX} {E}cosystem},}\ } (\bibinfo {year} {2020})\BibitemShut {NoStop}%
\bibitem [{\citenamefont {Devereux}\ \emph {et~al.}(2020)\citenamefont {Devereux}, \citenamefont {Smith}, \citenamefont {Huddleston}, \citenamefont {Barros}, \citenamefont {Zubatyuk}, \citenamefont {Isayev},\ and\ \citenamefont {Roitberg}}]{devereux2020extending}%
  \BibitemOpen
  \bibfield  {author} {\bibinfo {author} {\bibfnamefont {C.}~\bibnamefont {Devereux}}, \bibinfo {author} {\bibfnamefont {J.~S.}\ \bibnamefont {Smith}}, \bibinfo {author} {\bibfnamefont {K.~K.}\ \bibnamefont {Huddleston}}, \bibinfo {author} {\bibfnamefont {K.}~\bibnamefont {Barros}}, \bibinfo {author} {\bibfnamefont {R.}~\bibnamefont {Zubatyuk}}, \bibinfo {author} {\bibfnamefont {O.}~\bibnamefont {Isayev}}, \ and\ \bibinfo {author} {\bibfnamefont {A.~E.}\ \bibnamefont {Roitberg}},\ }\href@noop {} {\bibfield  {journal} {\bibinfo  {journal} {Journal of Chemical Theory and Computation}\ }\textbf {\bibinfo {volume} {16}},\ \bibinfo {pages} {4192} (\bibinfo {year} {2020})}\BibitemShut {NoStop}%
\bibitem [{\citenamefont {Jaffrelot~Inizan}\ \emph {et~al.}(2023)\citenamefont {Jaffrelot~Inizan}, \citenamefont {Plé}, \citenamefont {Adjoua}, \citenamefont {Ren}, \citenamefont {Gökcan}, \citenamefont {Isayev}, \citenamefont {Lagardère},\ and\ \citenamefont {Piquemal}}]{inizan2022scalable}%
  \BibitemOpen
  \bibfield  {author} {\bibinfo {author} {\bibfnamefont {T.}~\bibnamefont {Jaffrelot~Inizan}}, \bibinfo {author} {\bibfnamefont {T.}~\bibnamefont {Plé}}, \bibinfo {author} {\bibfnamefont {O.}~\bibnamefont {Adjoua}}, \bibinfo {author} {\bibfnamefont {P.}~\bibnamefont {Ren}}, \bibinfo {author} {\bibfnamefont {H.}~\bibnamefont {Gökcan}}, \bibinfo {author} {\bibfnamefont {O.}~\bibnamefont {Isayev}}, \bibinfo {author} {\bibfnamefont {L.}~\bibnamefont {Lagardère}}, \ and\ \bibinfo {author} {\bibfnamefont {J.-P.}\ \bibnamefont {Piquemal}},\ }\href {\doibase 10.1039/D2SC04815A} {\bibfield  {journal} {\bibinfo  {journal} {Chemical Science}\ }\textbf {\bibinfo {volume} {14}},\ \bibinfo {pages} {5438} (\bibinfo {year} {2023})}\BibitemShut {NoStop}%
\bibitem [{\citenamefont {Heek}\ \emph {et~al.}(2023)\citenamefont {Heek}, \citenamefont {Levskaya}, \citenamefont {Oliver}, \citenamefont {Ritter}, \citenamefont {Rondepierre}, \citenamefont {Steiner},\ and\ \citenamefont {van {Z}ee}}]{flax2020github}%
  \BibitemOpen
  \bibfield  {author} {\bibinfo {author} {\bibfnamefont {J.}~\bibnamefont {Heek}}, \bibinfo {author} {\bibfnamefont {A.}~\bibnamefont {Levskaya}}, \bibinfo {author} {\bibfnamefont {A.}~\bibnamefont {Oliver}}, \bibinfo {author} {\bibfnamefont {M.}~\bibnamefont {Ritter}}, \bibinfo {author} {\bibfnamefont {B.}~\bibnamefont {Rondepierre}}, \bibinfo {author} {\bibfnamefont {A.}~\bibnamefont {Steiner}}, \ and\ \bibinfo {author} {\bibfnamefont {M.}~\bibnamefont {van {Z}ee}},\ }\href {http://github.com/google/flax} {\enquote {\bibinfo {title} {{F}lax: A neural network library and ecosystem for {JAX}},}\ } (\bibinfo {year} {2023})\BibitemShut {NoStop}%
\bibitem [{\citenamefont {Ewald}(1921)}]{ewald1921ewald}%
  \BibitemOpen
  \bibfield  {author} {\bibinfo {author} {\bibfnamefont {P.~P.}\ \bibnamefont {Ewald}},\ }\href@noop {} {\bibfield  {journal} {\bibinfo  {journal} {Ann. Phys}\ }\textbf {\bibinfo {volume} {369}},\ \bibinfo {pages} {1} (\bibinfo {year} {1921})}\BibitemShut {NoStop}%
\bibitem [{\citenamefont {Behler}\ and\ \citenamefont {Parrinello}(2007)}]{behler2007generalized}%
  \BibitemOpen
  \bibfield  {author} {\bibinfo {author} {\bibfnamefont {J.}~\bibnamefont {Behler}}\ and\ \bibinfo {author} {\bibfnamefont {M.}~\bibnamefont {Parrinello}},\ }\href@noop {} {\bibfield  {journal} {\bibinfo  {journal} {Physical Review Letters}\ }\textbf {\bibinfo {volume} {98}},\ \bibinfo {pages} {146401} (\bibinfo {year} {2007})}\BibitemShut {NoStop}%
\bibitem [{\citenamefont {Gasteiger}, \citenamefont {Gro{\ss}},\ and\ \citenamefont {G{\"u}nnemann}(2020)}]{gasteiger2020directional}%
  \BibitemOpen
  \bibfield  {author} {\bibinfo {author} {\bibfnamefont {J.}~\bibnamefont {Gasteiger}}, \bibinfo {author} {\bibfnamefont {J.}~\bibnamefont {Gro{\ss}}}, \ and\ \bibinfo {author} {\bibfnamefont {S.}~\bibnamefont {G{\"u}nnemann}},\ }\href@noop {} {\bibfield  {journal} {\bibinfo  {journal} {arXiv preprint arXiv:2003.03123}\ } (\bibinfo {year} {2020})}\BibitemShut {NoStop}%
\bibitem [{\citenamefont {Rupp}\ \emph {et~al.}(2012)\citenamefont {Rupp}, \citenamefont {Tkatchenko}, \citenamefont {M{\"u}ller},\ and\ \citenamefont {Von~Lilienfeld}}]{rupp2012fast}%
  \BibitemOpen
  \bibfield  {author} {\bibinfo {author} {\bibfnamefont {M.}~\bibnamefont {Rupp}}, \bibinfo {author} {\bibfnamefont {A.}~\bibnamefont {Tkatchenko}}, \bibinfo {author} {\bibfnamefont {K.-R.}\ \bibnamefont {M{\"u}ller}}, \ and\ \bibinfo {author} {\bibfnamefont {O.~A.}\ \bibnamefont {Von~Lilienfeld}},\ }\href@noop {} {\bibfield  {journal} {\bibinfo  {journal} {Physical Review Letters}\ }\textbf {\bibinfo {volume} {108}},\ \bibinfo {pages} {058301} (\bibinfo {year} {2012})}\BibitemShut {NoStop}%
\bibitem [{\citenamefont {Eckhoff}\ and\ \citenamefont {Reiher}(2023)}]{eckhoff2023lifelong}%
  \BibitemOpen
  \bibfield  {author} {\bibinfo {author} {\bibfnamefont {M.}~\bibnamefont {Eckhoff}}\ and\ \bibinfo {author} {\bibfnamefont {M.}~\bibnamefont {Reiher}},\ }\href@noop {} {\bibfield  {journal} {\bibinfo  {journal} {Journal of Chemical Theory and Computation}\ }\textbf {\bibinfo {volume} {19}},\ \bibinfo {pages} {3509} (\bibinfo {year} {2023})}\BibitemShut {NoStop}%
\bibitem [{\citenamefont {Zhang}\ \emph {et~al.}(2018)\citenamefont {Zhang}, \citenamefont {Han}, \citenamefont {Wang}, \citenamefont {Saidi}, \citenamefont {Car} \emph {et~al.}}]{zhang2018end}%
  \BibitemOpen
  \bibfield  {author} {\bibinfo {author} {\bibfnamefont {L.}~\bibnamefont {Zhang}}, \bibinfo {author} {\bibfnamefont {J.}~\bibnamefont {Han}}, \bibinfo {author} {\bibfnamefont {H.}~\bibnamefont {Wang}}, \bibinfo {author} {\bibfnamefont {W.}~\bibnamefont {Saidi}}, \bibinfo {author} {\bibfnamefont {R.}~\bibnamefont {Car}},  \emph {et~al.},\ }\href@noop {} {\bibfield  {journal} {\bibinfo  {journal} {Advances in neural information processing systems}\ }\textbf {\bibinfo {volume} {31}} (\bibinfo {year} {2018})}\BibitemShut {NoStop}%
\bibitem [{\citenamefont {Sch{\"u}tt}, \citenamefont {Unke},\ and\ \citenamefont {Gastegger}(2021)}]{schutt2021equivariant}%
  \BibitemOpen
  \bibfield  {author} {\bibinfo {author} {\bibfnamefont {K.}~\bibnamefont {Sch{\"u}tt}}, \bibinfo {author} {\bibfnamefont {O.}~\bibnamefont {Unke}}, \ and\ \bibinfo {author} {\bibfnamefont {M.}~\bibnamefont {Gastegger}},\ }in\ \href@noop {} {\emph {\bibinfo {booktitle} {International Conference on Machine Learning}}}\ (\bibinfo {organization} {PMLR},\ \bibinfo {year} {2021})\ pp.\ \bibinfo {pages} {9377--9388}\BibitemShut {NoStop}%
\bibitem [{\citenamefont {Haghighatlari}\ \emph {et~al.}(2022)\citenamefont {Haghighatlari}, \citenamefont {Li}, \citenamefont {Guan}, \citenamefont {Zhang}, \citenamefont {Das}, \citenamefont {Stein}, \citenamefont {Heidar-Zadeh}, \citenamefont {Liu}, \citenamefont {Head-Gordon}, \citenamefont {Bertels} \emph {et~al.}}]{haghighatlari2022newtonnet}%
  \BibitemOpen
  \bibfield  {author} {\bibinfo {author} {\bibfnamefont {M.}~\bibnamefont {Haghighatlari}}, \bibinfo {author} {\bibfnamefont {J.}~\bibnamefont {Li}}, \bibinfo {author} {\bibfnamefont {X.}~\bibnamefont {Guan}}, \bibinfo {author} {\bibfnamefont {O.}~\bibnamefont {Zhang}}, \bibinfo {author} {\bibfnamefont {A.}~\bibnamefont {Das}}, \bibinfo {author} {\bibfnamefont {C.~J.}\ \bibnamefont {Stein}}, \bibinfo {author} {\bibfnamefont {F.}~\bibnamefont {Heidar-Zadeh}}, \bibinfo {author} {\bibfnamefont {M.}~\bibnamefont {Liu}}, \bibinfo {author} {\bibfnamefont {M.}~\bibnamefont {Head-Gordon}}, \bibinfo {author} {\bibfnamefont {L.}~\bibnamefont {Bertels}},  \emph {et~al.},\ }\href@noop {} {\bibfield  {journal} {\bibinfo  {journal} {Digital Discovery}\ }\textbf {\bibinfo {volume} {1}},\ \bibinfo {pages} {333} (\bibinfo {year} {2022})}\BibitemShut {NoStop}%
\bibitem [{\citenamefont {Geiger}\ and\ \citenamefont {Smidt}(2022)}]{geiger2022e3nn}%
  \BibitemOpen
  \bibfield  {author} {\bibinfo {author} {\bibfnamefont {M.}~\bibnamefont {Geiger}}\ and\ \bibinfo {author} {\bibfnamefont {T.}~\bibnamefont {Smidt}},\ }\href@noop {} {\bibfield  {journal} {\bibinfo  {journal} {arXiv preprint arXiv:2207.09453}\ } (\bibinfo {year} {2022})}\BibitemShut {NoStop}%
\bibitem [{\citenamefont {Kozinsky}\ \emph {et~al.}(2023)\citenamefont {Kozinsky}, \citenamefont {Musaelian}, \citenamefont {Johansson},\ and\ \citenamefont {Batzner}}]{kozinsky2023scaling}%
  \BibitemOpen
  \bibfield  {author} {\bibinfo {author} {\bibfnamefont {B.}~\bibnamefont {Kozinsky}}, \bibinfo {author} {\bibfnamefont {A.}~\bibnamefont {Musaelian}}, \bibinfo {author} {\bibfnamefont {A.}~\bibnamefont {Johansson}}, \ and\ \bibinfo {author} {\bibfnamefont {S.}~\bibnamefont {Batzner}},\ }in\ \href@noop {} {\emph {\bibinfo {booktitle} {Proceedings of the International Conference for High Performance Computing, Networking, Storage and Analysis}}}\ (\bibinfo {year} {2023})\ pp.\ \bibinfo {pages} {1--12}\BibitemShut {NoStop}%
\bibitem [{\citenamefont {Takamoto}, \citenamefont {Izumi},\ and\ \citenamefont {Li}(2022)}]{takamoto2022teanet}%
  \BibitemOpen
  \bibfield  {author} {\bibinfo {author} {\bibfnamefont {S.}~\bibnamefont {Takamoto}}, \bibinfo {author} {\bibfnamefont {S.}~\bibnamefont {Izumi}}, \ and\ \bibinfo {author} {\bibfnamefont {J.}~\bibnamefont {Li}},\ }\href@noop {} {\bibfield  {journal} {\bibinfo  {journal} {Computational Materials Science}\ }\textbf {\bibinfo {volume} {207}},\ \bibinfo {pages} {111280} (\bibinfo {year} {2022})}\BibitemShut {NoStop}%
\bibitem [{\citenamefont {Essmann}\ \emph {et~al.}(1995)\citenamefont {Essmann}, \citenamefont {Perera}, \citenamefont {Berkowitz}, \citenamefont {Darden}, \citenamefont {Lee},\ and\ \citenamefont {Pedersen}}]{essmann1995smooth}%
  \BibitemOpen
  \bibfield  {author} {\bibinfo {author} {\bibfnamefont {U.}~\bibnamefont {Essmann}}, \bibinfo {author} {\bibfnamefont {L.}~\bibnamefont {Perera}}, \bibinfo {author} {\bibfnamefont {M.~L.}\ \bibnamefont {Berkowitz}}, \bibinfo {author} {\bibfnamefont {T.}~\bibnamefont {Darden}}, \bibinfo {author} {\bibfnamefont {H.}~\bibnamefont {Lee}}, \ and\ \bibinfo {author} {\bibfnamefont {L.~G.}\ \bibnamefont {Pedersen}},\ }\href@noop {} {\bibfield  {journal} {\bibinfo  {journal} {The Journal of chemical physics}\ }\textbf {\bibinfo {volume} {103}},\ \bibinfo {pages} {8577} (\bibinfo {year} {1995})}\BibitemShut {NoStop}%
\bibitem [{\citenamefont {Khabibrakhmanov}, \citenamefont {Fedorov},\ and\ \citenamefont {Tkatchenko}(2023)}]{khabibrakhmanov2023universal}%
  \BibitemOpen
  \bibfield  {author} {\bibinfo {author} {\bibfnamefont {A.}~\bibnamefont {Khabibrakhmanov}}, \bibinfo {author} {\bibfnamefont {D.~V.}\ \bibnamefont {Fedorov}}, \ and\ \bibinfo {author} {\bibfnamefont {A.}~\bibnamefont {Tkatchenko}},\ }\href@noop {} {\bibfield  {journal} {\bibinfo  {journal} {Journal of Chemical Theory and Computation}\ }\textbf {\bibinfo {volume} {19}},\ \bibinfo {pages} {7895} (\bibinfo {year} {2023})}\BibitemShut {NoStop}%
\bibitem [{\citenamefont {Caldeweyher}\ \emph {et~al.}(2019)\citenamefont {Caldeweyher}, \citenamefont {Ehlert}, \citenamefont {Hansen}, \citenamefont {Neugebauer}, \citenamefont {Spicher}, \citenamefont {Bannwarth},\ and\ \citenamefont {Grimme}}]{caldeweyher2019generally}%
  \BibitemOpen
  \bibfield  {author} {\bibinfo {author} {\bibfnamefont {E.}~\bibnamefont {Caldeweyher}}, \bibinfo {author} {\bibfnamefont {S.}~\bibnamefont {Ehlert}}, \bibinfo {author} {\bibfnamefont {A.}~\bibnamefont {Hansen}}, \bibinfo {author} {\bibfnamefont {H.}~\bibnamefont {Neugebauer}}, \bibinfo {author} {\bibfnamefont {S.}~\bibnamefont {Spicher}}, \bibinfo {author} {\bibfnamefont {C.}~\bibnamefont {Bannwarth}}, \ and\ \bibinfo {author} {\bibfnamefont {S.}~\bibnamefont {Grimme}},\ }\href@noop {} {\bibfield  {journal} {\bibinfo  {journal} {The Journal of Chemical Physics}\ }\textbf {\bibinfo {volume} {150}} (\bibinfo {year} {2019})}\BibitemShut {NoStop}%
\bibitem [{\citenamefont {Ziegler}\ and\ \citenamefont {Biersack}(1985)}]{ziegler1985stopping}%
  \BibitemOpen
  \bibfield  {author} {\bibinfo {author} {\bibfnamefont {J.~F.}\ \bibnamefont {Ziegler}}\ and\ \bibinfo {author} {\bibfnamefont {J.~P.}\ \bibnamefont {Biersack}},\ }in\ \href@noop {} {\emph {\bibinfo {booktitle} {Treatise on heavy-ion science: volume 6: astrophysics, chemistry, and condensed matter}}}\ (\bibinfo  {publisher} {Springer},\ \bibinfo {year} {1985})\ pp.\ \bibinfo {pages} {93--129}\BibitemShut {NoStop}%
\bibitem [{\citenamefont {Hu}\ \emph {et~al.}(2021)\citenamefont {Hu}, \citenamefont {Shen}, \citenamefont {Wallis}, \citenamefont {Allen-Zhu}, \citenamefont {Li}, \citenamefont {Wang}, \citenamefont {Wang},\ and\ \citenamefont {Chen}}]{hu2021lora}%
  \BibitemOpen
  \bibfield  {author} {\bibinfo {author} {\bibfnamefont {E.~J.}\ \bibnamefont {Hu}}, \bibinfo {author} {\bibfnamefont {Y.}~\bibnamefont {Shen}}, \bibinfo {author} {\bibfnamefont {P.}~\bibnamefont {Wallis}}, \bibinfo {author} {\bibfnamefont {Z.}~\bibnamefont {Allen-Zhu}}, \bibinfo {author} {\bibfnamefont {Y.}~\bibnamefont {Li}}, \bibinfo {author} {\bibfnamefont {S.}~\bibnamefont {Wang}}, \bibinfo {author} {\bibfnamefont {L.}~\bibnamefont {Wang}}, \ and\ \bibinfo {author} {\bibfnamefont {W.}~\bibnamefont {Chen}},\ }\href@noop {} {\bibfield  {journal} {\bibinfo  {journal} {arXiv preprint arXiv:2106.09685}\ } (\bibinfo {year} {2021})}\BibitemShut {NoStop}%
\bibitem [{\citenamefont {Batatia}\ \emph {et~al.}(2022)\citenamefont {Batatia}, \citenamefont {Kovacs}, \citenamefont {Simm}, \citenamefont {Ortner},\ and\ \citenamefont {Cs{\'a}nyi}}]{batatia2022mace}%
  \BibitemOpen
  \bibfield  {author} {\bibinfo {author} {\bibfnamefont {I.}~\bibnamefont {Batatia}}, \bibinfo {author} {\bibfnamefont {D.~P.}\ \bibnamefont {Kovacs}}, \bibinfo {author} {\bibfnamefont {G.}~\bibnamefont {Simm}}, \bibinfo {author} {\bibfnamefont {C.}~\bibnamefont {Ortner}}, \ and\ \bibinfo {author} {\bibfnamefont {G.}~\bibnamefont {Cs{\'a}nyi}},\ }\href@noop {} {\bibfield  {journal} {\bibinfo  {journal} {Advances in Neural Information Processing Systems}\ }\textbf {\bibinfo {volume} {35}},\ \bibinfo {pages} {11423} (\bibinfo {year} {2022})}\BibitemShut {NoStop}%
\bibitem [{\citenamefont {Zhuang}\ \emph {et~al.}(2020)\citenamefont {Zhuang}, \citenamefont {Tang}, \citenamefont {Ding}, \citenamefont {Tatikonda}, \citenamefont {Dvornek}, \citenamefont {Papademetris},\ and\ \citenamefont {Duncan}}]{zhuang2020adabelief}%
  \BibitemOpen
  \bibfield  {author} {\bibinfo {author} {\bibfnamefont {J.}~\bibnamefont {Zhuang}}, \bibinfo {author} {\bibfnamefont {T.}~\bibnamefont {Tang}}, \bibinfo {author} {\bibfnamefont {Y.}~\bibnamefont {Ding}}, \bibinfo {author} {\bibfnamefont {S.~C.}\ \bibnamefont {Tatikonda}}, \bibinfo {author} {\bibfnamefont {N.}~\bibnamefont {Dvornek}}, \bibinfo {author} {\bibfnamefont {X.}~\bibnamefont {Papademetris}}, \ and\ \bibinfo {author} {\bibfnamefont {J.}~\bibnamefont {Duncan}},\ }\href@noop {} {\bibfield  {journal} {\bibinfo  {journal} {Advances in neural information processing systems}\ }\textbf {\bibinfo {volume} {33}},\ \bibinfo {pages} {18795} (\bibinfo {year} {2020})}\BibitemShut {NoStop}%
\bibitem [{\citenamefont {Loshchilov}\ and\ \citenamefont {Hutter}(2017)}]{loshchilov2017decoupled}%
  \BibitemOpen
  \bibfield  {author} {\bibinfo {author} {\bibfnamefont {I.}~\bibnamefont {Loshchilov}}\ and\ \bibinfo {author} {\bibfnamefont {F.}~\bibnamefont {Hutter}},\ }\href@noop {} {\bibfield  {journal} {\bibinfo  {journal} {arXiv preprint arXiv:1711.05101}\ } (\bibinfo {year} {2017})}\BibitemShut {NoStop}%
\bibitem [{\citenamefont {Brock}\ \emph {et~al.}(2021)\citenamefont {Brock}, \citenamefont {De}, \citenamefont {Smith},\ and\ \citenamefont {Simonyan}}]{brock2021high}%
  \BibitemOpen
  \bibfield  {author} {\bibinfo {author} {\bibfnamefont {A.}~\bibnamefont {Brock}}, \bibinfo {author} {\bibfnamefont {S.}~\bibnamefont {De}}, \bibinfo {author} {\bibfnamefont {S.~L.}\ \bibnamefont {Smith}}, \ and\ \bibinfo {author} {\bibfnamefont {K.}~\bibnamefont {Simonyan}},\ }in\ \href@noop {} {\emph {\bibinfo {booktitle} {International Conference on Machine Learning}}}\ (\bibinfo {organization} {PMLR},\ \bibinfo {year} {2021})\ pp.\ \bibinfo {pages} {1059--1071}\BibitemShut {NoStop}%
\bibitem [{\citenamefont {Smith}\ and\ \citenamefont {Topin}(2019)}]{smith2019super}%
  \BibitemOpen
  \bibfield  {author} {\bibinfo {author} {\bibfnamefont {L.~N.}\ \bibnamefont {Smith}}\ and\ \bibinfo {author} {\bibfnamefont {N.}~\bibnamefont {Topin}},\ }in\ \href@noop {} {\emph {\bibinfo {booktitle} {Artificial intelligence and machine learning for multi-domain operations applications}}},\ Vol.\ \bibinfo {volume} {11006}\ (\bibinfo {organization} {SPIE},\ \bibinfo {year} {2019})\ pp.\ \bibinfo {pages} {369--386}\BibitemShut {NoStop}%
\bibitem [{\citenamefont {Kellner}\ and\ \citenamefont {Ceriotti}(2024)}]{kellner2024uncertainty}%
  \BibitemOpen
  \bibfield  {author} {\bibinfo {author} {\bibfnamefont {M.}~\bibnamefont {Kellner}}\ and\ \bibinfo {author} {\bibfnamefont {M.}~\bibnamefont {Ceriotti}},\ }\href@noop {} {\bibfield  {journal} {\bibinfo  {journal} {arXiv preprint arXiv:2402.16621}\ } (\bibinfo {year} {2024})}\BibitemShut {NoStop}%
\bibitem [{\citenamefont {Amini}\ \emph {et~al.}(2020)\citenamefont {Amini}, \citenamefont {Schwarting}, \citenamefont {Soleimany},\ and\ \citenamefont {Rus}}]{amini2020deep}%
  \BibitemOpen
  \bibfield  {author} {\bibinfo {author} {\bibfnamefont {A.}~\bibnamefont {Amini}}, \bibinfo {author} {\bibfnamefont {W.}~\bibnamefont {Schwarting}}, \bibinfo {author} {\bibfnamefont {A.}~\bibnamefont {Soleimany}}, \ and\ \bibinfo {author} {\bibfnamefont {D.}~\bibnamefont {Rus}},\ }\href@noop {} {\bibfield  {journal} {\bibinfo  {journal} {Advances in neural information processing systems}\ }\textbf {\bibinfo {volume} {33}},\ \bibinfo {pages} {14927} (\bibinfo {year} {2020})}\BibitemShut {NoStop}%
\bibitem [{\citenamefont {Meinert}\ and\ \citenamefont {Lavin}(2021)}]{meinert2021multivariate}%
  \BibitemOpen
  \bibfield  {author} {\bibinfo {author} {\bibfnamefont {N.}~\bibnamefont {Meinert}}\ and\ \bibinfo {author} {\bibfnamefont {A.}~\bibnamefont {Lavin}},\ }\href@noop {} {\bibfield  {journal} {\bibinfo  {journal} {arXiv preprint arXiv:2104.06135}\ } (\bibinfo {year} {2021})}\BibitemShut {NoStop}%
\bibitem [{\citenamefont {Larsen}\ \emph {et~al.}(2017)\citenamefont {Larsen}, \citenamefont {Mortensen}, \citenamefont {Blomqvist}, \citenamefont {Castelli}, \citenamefont {Christensen}, \citenamefont {Du{\l}ak}, \citenamefont {Friis}, \citenamefont {Groves}, \citenamefont {Hammer}, \citenamefont {Hargus} \emph {et~al.}}]{larsen2017atomic}%
  \BibitemOpen
  \bibfield  {author} {\bibinfo {author} {\bibfnamefont {A.~H.}\ \bibnamefont {Larsen}}, \bibinfo {author} {\bibfnamefont {J.~J.}\ \bibnamefont {Mortensen}}, \bibinfo {author} {\bibfnamefont {J.}~\bibnamefont {Blomqvist}}, \bibinfo {author} {\bibfnamefont {I.~E.}\ \bibnamefont {Castelli}}, \bibinfo {author} {\bibfnamefont {R.}~\bibnamefont {Christensen}}, \bibinfo {author} {\bibfnamefont {M.}~\bibnamefont {Du{\l}ak}}, \bibinfo {author} {\bibfnamefont {J.}~\bibnamefont {Friis}}, \bibinfo {author} {\bibfnamefont {M.~N.}\ \bibnamefont {Groves}}, \bibinfo {author} {\bibfnamefont {B.}~\bibnamefont {Hammer}}, \bibinfo {author} {\bibfnamefont {C.}~\bibnamefont {Hargus}},  \emph {et~al.},\ }\href@noop {} {\bibfield  {journal} {\bibinfo  {journal} {Journal of Physics: Condensed Matter}\ }\textbf {\bibinfo {volume} {29}},\ \bibinfo {pages} {273002} (\bibinfo {year} {2017})}\BibitemShut {NoStop}%
\bibitem [{\citenamefont {Kapil}\ \emph {et~al.}(2019)\citenamefont {Kapil}, \citenamefont {Rossi}, \citenamefont {Marsalek}, \citenamefont {Petraglia}, \citenamefont {Litman}, \citenamefont {Spura}, \citenamefont {Cheng}, \citenamefont {Cuzzocrea}, \citenamefont {Mei{\ss}ner}, \citenamefont {Wilkins} \emph {et~al.}}]{kapil2019pi}%
  \BibitemOpen
  \bibfield  {author} {\bibinfo {author} {\bibfnamefont {V.}~\bibnamefont {Kapil}}, \bibinfo {author} {\bibfnamefont {M.}~\bibnamefont {Rossi}}, \bibinfo {author} {\bibfnamefont {O.}~\bibnamefont {Marsalek}}, \bibinfo {author} {\bibfnamefont {R.}~\bibnamefont {Petraglia}}, \bibinfo {author} {\bibfnamefont {Y.}~\bibnamefont {Litman}}, \bibinfo {author} {\bibfnamefont {T.}~\bibnamefont {Spura}}, \bibinfo {author} {\bibfnamefont {B.}~\bibnamefont {Cheng}}, \bibinfo {author} {\bibfnamefont {A.}~\bibnamefont {Cuzzocrea}}, \bibinfo {author} {\bibfnamefont {R.~H.}\ \bibnamefont {Mei{\ss}ner}}, \bibinfo {author} {\bibfnamefont {D.~M.}\ \bibnamefont {Wilkins}},  \emph {et~al.},\ }\href@noop {} {\bibfield  {journal} {\bibinfo  {journal} {Computer Physics Communications}\ }\textbf {\bibinfo {volume} {236}},\ \bibinfo {pages} {214} (\bibinfo {year} {2019})}\BibitemShut {NoStop}%
\bibitem [{\citenamefont {Plé}\ \emph {et~al.}(2023)\citenamefont {Plé}, \citenamefont {Mauger}, \citenamefont {Adjoua}, \citenamefont {Inizan}, \citenamefont {Lagardère}, \citenamefont {Huppert},\ and\ \citenamefont {Piquemal}}]{ple2022routine}%
  \BibitemOpen
  \bibfield  {author} {\bibinfo {author} {\bibfnamefont {T.}~\bibnamefont {Plé}}, \bibinfo {author} {\bibfnamefont {N.}~\bibnamefont {Mauger}}, \bibinfo {author} {\bibfnamefont {O.}~\bibnamefont {Adjoua}}, \bibinfo {author} {\bibfnamefont {T.~J.}\ \bibnamefont {Inizan}}, \bibinfo {author} {\bibfnamefont {L.}~\bibnamefont {Lagardère}}, \bibinfo {author} {\bibfnamefont {S.}~\bibnamefont {Huppert}}, \ and\ \bibinfo {author} {\bibfnamefont {J.-P.}\ \bibnamefont {Piquemal}},\ }\href {\doibase 10.1021/acs.jctc.2c01233} {\bibfield  {journal} {\bibinfo  {journal} {Journal of Chemical Theory and Computation}\ }\textbf {\bibinfo {volume} {19}},\ \bibinfo {pages} {1432} (\bibinfo {year} {2023})},\ \bibinfo {note} {pMID: 36856658},\ \Eprint {http://arxiv.org/abs/https://doi.org/10.1021/acs.jctc.2c01233} {https://doi.org/10.1021/acs.jctc.2c01233} \BibitemShut {NoStop}%
\bibitem [{\citenamefont {Lahey}\ and\ \citenamefont {Rowley}(2020)}]{lahey2020simulating}%
  \BibitemOpen
  \bibfield  {author} {\bibinfo {author} {\bibfnamefont {S.-L.~J.}\ \bibnamefont {Lahey}}\ and\ \bibinfo {author} {\bibfnamefont {C.~N.}\ \bibnamefont {Rowley}},\ }\href@noop {} {\bibfield  {journal} {\bibinfo  {journal} {Chemical Science}\ }\textbf {\bibinfo {volume} {11}},\ \bibinfo {pages} {2362} (\bibinfo {year} {2020})}\BibitemShut {NoStop}%
\bibitem [{\citenamefont {Galvelis}\ \emph {et~al.}(2023)\citenamefont {Galvelis}, \citenamefont {Varela-Rial}, \citenamefont {Doerr}, \citenamefont {Fino}, \citenamefont {Eastman}, \citenamefont {Markland}, \citenamefont {Chodera},\ and\ \citenamefont {De~Fabritiis}}]{galvelis2023nnp}%
  \BibitemOpen
  \bibfield  {author} {\bibinfo {author} {\bibfnamefont {R.}~\bibnamefont {Galvelis}}, \bibinfo {author} {\bibfnamefont {A.}~\bibnamefont {Varela-Rial}}, \bibinfo {author} {\bibfnamefont {S.}~\bibnamefont {Doerr}}, \bibinfo {author} {\bibfnamefont {R.}~\bibnamefont {Fino}}, \bibinfo {author} {\bibfnamefont {P.}~\bibnamefont {Eastman}}, \bibinfo {author} {\bibfnamefont {T.~E.}\ \bibnamefont {Markland}}, \bibinfo {author} {\bibfnamefont {J.~D.}\ \bibnamefont {Chodera}}, \ and\ \bibinfo {author} {\bibfnamefont {G.}~\bibnamefont {De~Fabritiis}},\ }\href@noop {} {\bibfield  {journal} {\bibinfo  {journal} {Journal of Chemical Information and Modeling}\ }\textbf {\bibinfo {volume} {63}},\ \bibinfo {pages} {5701} (\bibinfo {year} {2023})}\BibitemShut {NoStop}%
\bibitem [{\citenamefont {Illarionov}\ \emph {et~al.}(2023)\citenamefont {Illarionov}, \citenamefont {Sakipov}, \citenamefont {Pereyaslavets}, \citenamefont {Kurnikov}, \citenamefont {Kamath}, \citenamefont {Butin}, \citenamefont {Voronina}, \citenamefont {Ivahnenko}, \citenamefont {Leontyev}, \citenamefont {Nawrocki} \emph {et~al.}}]{illarionov2023combining}%
  \BibitemOpen
  \bibfield  {author} {\bibinfo {author} {\bibfnamefont {A.}~\bibnamefont {Illarionov}}, \bibinfo {author} {\bibfnamefont {S.}~\bibnamefont {Sakipov}}, \bibinfo {author} {\bibfnamefont {L.}~\bibnamefont {Pereyaslavets}}, \bibinfo {author} {\bibfnamefont {I.~V.}\ \bibnamefont {Kurnikov}}, \bibinfo {author} {\bibfnamefont {G.}~\bibnamefont {Kamath}}, \bibinfo {author} {\bibfnamefont {O.}~\bibnamefont {Butin}}, \bibinfo {author} {\bibfnamefont {E.}~\bibnamefont {Voronina}}, \bibinfo {author} {\bibfnamefont {I.}~\bibnamefont {Ivahnenko}}, \bibinfo {author} {\bibfnamefont {I.}~\bibnamefont {Leontyev}}, \bibinfo {author} {\bibfnamefont {G.}~\bibnamefont {Nawrocki}},  \emph {et~al.},\ }\href@noop {} {\bibfield  {journal} {\bibinfo  {journal} {Journal of the American Chemical Society}\ }\textbf {\bibinfo {volume} {145}},\ \bibinfo {pages} {23620} (\bibinfo {year} {2023})}\BibitemShut {NoStop}%
\bibitem [{\citenamefont {Wang}\ \emph {et~al.}(2024{\natexlab{b}})\citenamefont {Wang}, \citenamefont {Inizan}, \citenamefont {Liu}, \citenamefont {Piquemal},\ and\ \citenamefont {Ren}}]{wang2024incorporating}%
  \BibitemOpen
  \bibfield  {author} {\bibinfo {author} {\bibfnamefont {Y.}~\bibnamefont {Wang}}, \bibinfo {author} {\bibfnamefont {T.~J.}\ \bibnamefont {Inizan}}, \bibinfo {author} {\bibfnamefont {C.}~\bibnamefont {Liu}}, \bibinfo {author} {\bibfnamefont {J.-P.}\ \bibnamefont {Piquemal}}, \ and\ \bibinfo {author} {\bibfnamefont {P.}~\bibnamefont {Ren}},\ }\href@noop {} {\bibfield  {journal} {\bibinfo  {journal} {The Journal of Physical Chemistry B}\ }\textbf {\bibinfo {volume} {128}},\ \bibinfo {pages} {2381} (\bibinfo {year} {2024}{\natexlab{b}})}\BibitemShut {NoStop}%
\bibitem [{\citenamefont {Leimkuhler}\ and\ \citenamefont {Matthews}(2013)}]{leimkuhler2013rational}%
  \BibitemOpen
  \bibfield  {author} {\bibinfo {author} {\bibfnamefont {B.}~\bibnamefont {Leimkuhler}}\ and\ \bibinfo {author} {\bibfnamefont {C.}~\bibnamefont {Matthews}},\ }\href@noop {} {\bibfield  {journal} {\bibinfo  {journal} {Applied Mathematics Research eXpress}\ }\textbf {\bibinfo {volume} {2013}},\ \bibinfo {pages} {34} (\bibinfo {year} {2013})}\BibitemShut {NoStop}%
\bibitem [{\citenamefont {Mangaud}\ \emph {et~al.}(2019)\citenamefont {Mangaud}, \citenamefont {Huppert}, \citenamefont {Pl{\'e}}, \citenamefont {Depondt}, \citenamefont {Bonella},\ and\ \citenamefont {Finocchi}}]{mangaud2019fluctuation}%
  \BibitemOpen
  \bibfield  {author} {\bibinfo {author} {\bibfnamefont {E.}~\bibnamefont {Mangaud}}, \bibinfo {author} {\bibfnamefont {S.}~\bibnamefont {Huppert}}, \bibinfo {author} {\bibfnamefont {T.}~\bibnamefont {Pl{\'e}}}, \bibinfo {author} {\bibfnamefont {P.}~\bibnamefont {Depondt}}, \bibinfo {author} {\bibfnamefont {S.}~\bibnamefont {Bonella}}, \ and\ \bibinfo {author} {\bibfnamefont {F.}~\bibnamefont {Finocchi}},\ }\href@noop {} {\bibfield  {journal} {\bibinfo  {journal} {Journal of Chemical Theory and Computation}\ }\textbf {\bibinfo {volume} {15}},\ \bibinfo {pages} {2863} (\bibinfo {year} {2019})}\BibitemShut {NoStop}%
\bibitem [{\citenamefont {Habershon}\ \emph {et~al.}(2013)\citenamefont {Habershon}, \citenamefont {Manolopoulos}, \citenamefont {Markland},\ and\ \citenamefont {Miller~III}}]{habershon2013ring}%
  \BibitemOpen
  \bibfield  {author} {\bibinfo {author} {\bibfnamefont {S.}~\bibnamefont {Habershon}}, \bibinfo {author} {\bibfnamefont {D.~E.}\ \bibnamefont {Manolopoulos}}, \bibinfo {author} {\bibfnamefont {T.~E.}\ \bibnamefont {Markland}}, \ and\ \bibinfo {author} {\bibfnamefont {T.~F.}\ \bibnamefont {Miller~III}},\ }\href@noop {} {\bibfield  {journal} {\bibinfo  {journal} {Annual Review of Physical Chemistry}\ }\textbf {\bibinfo {volume} {64}},\ \bibinfo {pages} {387} (\bibinfo {year} {2013})}\BibitemShut {NoStop}%
\bibitem [{\citenamefont {Mauger}\ \emph {et~al.}(2021)\citenamefont {Mauger}, \citenamefont {Plé}, \citenamefont {Lagardère}, \citenamefont {Bonella}, \citenamefont {Mangaud}, \citenamefont {Piquemal},\ and\ \citenamefont {Huppert}}]{mauger2021nuclear}%
  \BibitemOpen
  \bibfield  {author} {\bibinfo {author} {\bibfnamefont {N.}~\bibnamefont {Mauger}}, \bibinfo {author} {\bibfnamefont {T.}~\bibnamefont {Plé}}, \bibinfo {author} {\bibfnamefont {L.}~\bibnamefont {Lagardère}}, \bibinfo {author} {\bibfnamefont {S.}~\bibnamefont {Bonella}}, \bibinfo {author} {\bibfnamefont {E.}~\bibnamefont {Mangaud}}, \bibinfo {author} {\bibfnamefont {J.-P.}\ \bibnamefont {Piquemal}}, \ and\ \bibinfo {author} {\bibfnamefont {S.}~\bibnamefont {Huppert}},\ }\href {\doibase 10.1021/acs.jpclett.1c01722} {\bibfield  {journal} {\bibinfo  {journal} {The Journal of Physical Chemistry Letters}\ }\textbf {\bibinfo {volume} {12}},\ \bibinfo {pages} {8285} (\bibinfo {year} {2021})},\ \bibinfo {note} {pMID: 34427440},\ \Eprint {http://arxiv.org/abs/https://doi.org/10.1021/acs.jpclett.1c01722} {https://doi.org/10.1021/acs.jpclett.1c01722} \BibitemShut {NoStop}%
\bibitem [{\citenamefont {Gao}\ \emph {et~al.}(2020)\citenamefont {Gao}, \citenamefont {Ramezanghorbani}, \citenamefont {Isayev}, \citenamefont {Smith},\ and\ \citenamefont {Roitberg}}]{gao2020torchani}%
  \BibitemOpen
  \bibfield  {author} {\bibinfo {author} {\bibfnamefont {X.}~\bibnamefont {Gao}}, \bibinfo {author} {\bibfnamefont {F.}~\bibnamefont {Ramezanghorbani}}, \bibinfo {author} {\bibfnamefont {O.}~\bibnamefont {Isayev}}, \bibinfo {author} {\bibfnamefont {J.~S.}\ \bibnamefont {Smith}}, \ and\ \bibinfo {author} {\bibfnamefont {A.~E.}\ \bibnamefont {Roitberg}},\ }\href@noop {} {\bibfield  {journal} {\bibinfo  {journal} {Journal of Chemical Information and Modeling}\ }\textbf {\bibinfo {volume} {60}},\ \bibinfo {pages} {3408} (\bibinfo {year} {2020})}\BibitemShut {NoStop}%
\bibitem [{\citenamefont {Lagardere}, \citenamefont {Aviat},\ and\ \citenamefont {Piquemal}(2019)}]{lagardere2019pushing}%
  \BibitemOpen
  \bibfield  {author} {\bibinfo {author} {\bibfnamefont {L.}~\bibnamefont {Lagardere}}, \bibinfo {author} {\bibfnamefont {F.}~\bibnamefont {Aviat}}, \ and\ \bibinfo {author} {\bibfnamefont {J.-P.}\ \bibnamefont {Piquemal}},\ }\href@noop {} {\bibfield  {journal} {\bibinfo  {journal} {The journal of physical chemistry letters}\ }\textbf {\bibinfo {volume} {10}},\ \bibinfo {pages} {2593} (\bibinfo {year} {2019})}\BibitemShut {NoStop}%
\bibitem [{\citenamefont {Hoja}\ \emph {et~al.}(2021)\citenamefont {Hoja}, \citenamefont {Medrano~Sandonas}, \citenamefont {Ernst}, \citenamefont {Vazquez-Mayagoitia}, \citenamefont {DiStasio~Jr},\ and\ \citenamefont {Tkatchenko}}]{hoja2021qm7}%
  \BibitemOpen
  \bibfield  {author} {\bibinfo {author} {\bibfnamefont {J.}~\bibnamefont {Hoja}}, \bibinfo {author} {\bibfnamefont {L.}~\bibnamefont {Medrano~Sandonas}}, \bibinfo {author} {\bibfnamefont {B.~G.}\ \bibnamefont {Ernst}}, \bibinfo {author} {\bibfnamefont {A.}~\bibnamefont {Vazquez-Mayagoitia}}, \bibinfo {author} {\bibfnamefont {R.~A.}\ \bibnamefont {DiStasio~Jr}}, \ and\ \bibinfo {author} {\bibfnamefont {A.}~\bibnamefont {Tkatchenko}},\ }\href@noop {} {\bibfield  {journal} {\bibinfo  {journal} {Scientific data}\ }\textbf {\bibinfo {volume} {8}},\ \bibinfo {pages} {43} (\bibinfo {year} {2021})}\BibitemShut {NoStop}%
\bibitem [{\citenamefont {Smith}\ \emph {et~al.}(2020)\citenamefont {Smith}, \citenamefont {Zubatyuk}, \citenamefont {Nebgen}, \citenamefont {Lubbers}, \citenamefont {Barros}, \citenamefont {Roitberg}, \citenamefont {Isayev},\ and\ \citenamefont {Tretiak}}]{smith2020ani}%
  \BibitemOpen
  \bibfield  {author} {\bibinfo {author} {\bibfnamefont {J.~S.}\ \bibnamefont {Smith}}, \bibinfo {author} {\bibfnamefont {R.}~\bibnamefont {Zubatyuk}}, \bibinfo {author} {\bibfnamefont {B.}~\bibnamefont {Nebgen}}, \bibinfo {author} {\bibfnamefont {N.}~\bibnamefont {Lubbers}}, \bibinfo {author} {\bibfnamefont {K.}~\bibnamefont {Barros}}, \bibinfo {author} {\bibfnamefont {A.~E.}\ \bibnamefont {Roitberg}}, \bibinfo {author} {\bibfnamefont {O.}~\bibnamefont {Isayev}}, \ and\ \bibinfo {author} {\bibfnamefont {S.}~\bibnamefont {Tretiak}},\ }\href@noop {} {\bibfield  {journal} {\bibinfo  {journal} {Scientific data}\ }\textbf {\bibinfo {volume} {7}},\ \bibinfo {pages} {134} (\bibinfo {year} {2020})}\BibitemShut {NoStop}%
\bibitem [{\citenamefont {Eastman}\ \emph {et~al.}(2023)\citenamefont {Eastman}, \citenamefont {Behara}, \citenamefont {Dotson}, \citenamefont {Galvelis}, \citenamefont {Herr}, \citenamefont {Horton}, \citenamefont {Mao}, \citenamefont {Chodera}, \citenamefont {Pritchard}, \citenamefont {Wang} \emph {et~al.}}]{eastman2023spice}%
  \BibitemOpen
  \bibfield  {author} {\bibinfo {author} {\bibfnamefont {P.}~\bibnamefont {Eastman}}, \bibinfo {author} {\bibfnamefont {P.~K.}\ \bibnamefont {Behara}}, \bibinfo {author} {\bibfnamefont {D.~L.}\ \bibnamefont {Dotson}}, \bibinfo {author} {\bibfnamefont {R.}~\bibnamefont {Galvelis}}, \bibinfo {author} {\bibfnamefont {J.~E.}\ \bibnamefont {Herr}}, \bibinfo {author} {\bibfnamefont {J.~T.}\ \bibnamefont {Horton}}, \bibinfo {author} {\bibfnamefont {Y.}~\bibnamefont {Mao}}, \bibinfo {author} {\bibfnamefont {J.~D.}\ \bibnamefont {Chodera}}, \bibinfo {author} {\bibfnamefont {B.~P.}\ \bibnamefont {Pritchard}}, \bibinfo {author} {\bibfnamefont {Y.}~\bibnamefont {Wang}},  \emph {et~al.},\ }\href@noop {} {\bibfield  {journal} {\bibinfo  {journal} {Scientific Data}\ }\textbf {\bibinfo {volume} {10}},\ \bibinfo {pages} {11} (\bibinfo {year} {2023})}\BibitemShut {NoStop}%
\bibitem [{\citenamefont {Verstraelen}\ \emph {et~al.}(2016)\citenamefont {Verstraelen}, \citenamefont {Vandenbrande}, \citenamefont {Heidar-Zadeh}, \citenamefont {Vanduyfhuys}, \citenamefont {Van~Speybroeck}, \citenamefont {Waroquier},\ and\ \citenamefont {Ayers}}]{verstraelen2016minimal}%
  \BibitemOpen
  \bibfield  {author} {\bibinfo {author} {\bibfnamefont {T.}~\bibnamefont {Verstraelen}}, \bibinfo {author} {\bibfnamefont {S.}~\bibnamefont {Vandenbrande}}, \bibinfo {author} {\bibfnamefont {F.}~\bibnamefont {Heidar-Zadeh}}, \bibinfo {author} {\bibfnamefont {L.}~\bibnamefont {Vanduyfhuys}}, \bibinfo {author} {\bibfnamefont {V.}~\bibnamefont {Van~Speybroeck}}, \bibinfo {author} {\bibfnamefont {M.}~\bibnamefont {Waroquier}}, \ and\ \bibinfo {author} {\bibfnamefont {P.~W.}\ \bibnamefont {Ayers}},\ }\href@noop {} {\bibfield  {journal} {\bibinfo  {journal} {Journal of Chemical Theory and Computation}\ }\textbf {\bibinfo {volume} {12}},\ \bibinfo {pages} {3894} (\bibinfo {year} {2016})}\BibitemShut {NoStop}%
\bibitem [{\citenamefont {Zubatyuk}\ \emph {et~al.}(2021)\citenamefont {Zubatyuk}, \citenamefont {Smith}, \citenamefont {Nebgen}, \citenamefont {Tretiak},\ and\ \citenamefont {Isayev}}]{zubatyuk2021teaching}%
  \BibitemOpen
  \bibfield  {author} {\bibinfo {author} {\bibfnamefont {R.}~\bibnamefont {Zubatyuk}}, \bibinfo {author} {\bibfnamefont {J.~S.}\ \bibnamefont {Smith}}, \bibinfo {author} {\bibfnamefont {B.~T.}\ \bibnamefont {Nebgen}}, \bibinfo {author} {\bibfnamefont {S.}~\bibnamefont {Tretiak}}, \ and\ \bibinfo {author} {\bibfnamefont {O.}~\bibnamefont {Isayev}},\ }\href@noop {} {\bibfield  {journal} {\bibinfo  {journal} {Nature Communications}\ }\textbf {\bibinfo {volume} {12}},\ \bibinfo {pages} {1} (\bibinfo {year} {2021})}\BibitemShut {NoStop}%
\end{thebibliography}%

\end{document}